\documentclass[aps,twocolumn,superscriptaddress,showpacs]{revtex4}

\usepackage[pdftex]{graphicx}
\usepackage{bm}
\usepackage{natbib}

\usepackage{amsmath}

\usepackage[english]{babel}

\begin{document}

\title{Human Activity in the Web}

\author{Filippo Radicchi\footnote{Correspondence should be addressed to f.radicchi@gmail.com}}
\affiliation{Complex Networks \& Systems, ISI Foundation, 10133 Turin, Italy}

\begin{abstract}
The recent information technology revolution has enabled the analysis and processing of large-scale datasets describing human activities. The main source of data is represented by the Web, where  humans generally use to spend a relevant part of their day. Here we study three large datasets containing the information about Web human activities in different contexts. We study in details inter-event and waiting time statistics. In both cases, the number of subsequent operations which differ by $\tau$ units of time decays power-like as $\tau$ increases. We use non-parametric statistical tests in order to estimate the significance level of reliability of global distributions to describe activity patterns of single users. Global inter-event time probability distributions are not representative for the behavior of single users: the shape of single users'inter-event distributions is strongly influenced by the total number of operations performed by the users and distributions of the total number of operations performed by users are heterogeneous. A universal behavior can be anyway found by suppressing the intrinsic dependence of the global probability distribution on the activity of the users. This suppression can be performed by simply dividing the inter-event times with their average values. Differently, waiting time probability distributions seem to be independent of the activity of users and global probability distributions are able to significantly represent the replying activity patterns of single users.
\end{abstract}


\maketitle

\section{Introduction}
Recent years have evidenced a great interest in understanding and modeling human behavior~\cite{castellano07}. The scientific attention to this topic is motivated by clear economic and technological purposes since the possibility to monitor  and mathematically describe human behavior  may have important implications in  resource management and service allocation.  Examples of empirically studied human activities range from communication patterns of e-mails~\cite{ebel02,  eckmann04, johansen04, barabasi05, vazquez06, malmgren08} and surface mails~\cite{oliveira05} to Web surfing~\cite{johansen01, dezso06, vazquez06, goncalves08}, from printing requests~\cite{hardera06} to library loans~\cite{vazquez06}. The main result, arising from all these studies, concerns the bursty behavior of humans~\cite{barabasi05}: the time difference (namely $\tau$) between two consecutive human actions follow a power-law distribution [i.e., $P\left(\tau\right) \sim \tau^{-\beta}$]. The burstiness of humans therefore consists of long periods of inactivity followed by short periods of time in which humans concentrate their actions.

\

In this paper we take the advantage of very large datasets describing human activities in the Web.  Differently from former studies, our data describe activities which are not necessarily related with daily routines, as for example sending and receiving e-mails: two consecutive actions performed by the same person may differ of an amount of time of the order of days, weeks, months and even years. The nature of our datasets allows therefore the statistical study of inter-event and waiting time probability distribution functions (pdf) defined over a wide range of possible values, where the time gaps between two consecutive actions of the same user may be even longer than one year. Interestingly, the results show a clear bursty behavior of human activity over the whole range of possible values.  We provide a statistical non-parametric test able to quantify the reliability of the global inter-event and waiting time pdfs (global in the sense that they are calculated over all users) in order to predict the same distributions in the case of single users. For inter-event time pdfs, we find that the decay exponents strongly depend on the activity of the users~\cite{zhou08} and therefore pdfs corresponding to different level of activity are more representative than a global one. This finding suggests to suppress the dependence of the inter-event time by considering relative quantities instead of absolute ones. If the variables representing the inter-event times are divided by their average values, the new variables obey, independently on the activity of single users, the same distribution and the single users' pdfs  are well represented by the global pdf. Differently, in the case of the waiting time pdfs the decay exponents do not depend on the activity of the users and the global pdf well describes the activity patterns of single users.  

\

The paper is organized as follows. In section~\ref{sec:data}, we give a detailed description of the data used in our empirical analysis. In section~\ref{sec:act}, we show that populations of users present an heterogeneous degree of activity. We then start to consider inter-event and waiting time statistics (sections~\ref{sec:time} and~\ref{sec:wait}). In section~\ref{sec:time_a}, we compute the global inter-event time pdfs and we characterize them by estimating the decay exponents. In section~\ref{sec:time_b}, we statistically test the reliability of the inter-event time pdfs to describe the real activity of single users. Since the activity patterns of single users are in general not well described by the global inter-event time pdf, we calculate the inter-event time pdfs for users who have performed a similar number of actions and show that these distributions (i) well describe the activity patterns of single users and (ii) are in general different each other. The previous results suggest the possibility to find a more general rule. In section~\ref{sec:scaling}, we suppress the dependence on the number of activities of the variables representing the inter-event times by simply dividing these quantities with their average values. The new variables generate new single users' pdfs: the global pdf of rescaled inter-event times is able to significantly describe the activity patterns of single users. In section~\ref{sec:wait}, we calculate the time gap between messages and their replies (waiting times) and the statistics associated with them. In this case,  the global pdf is able to significantly describe the behavior of single users. In section~\ref{sec:end} we summarize the results of the paper and formulate our final considerations.

\section{Datasets description}
\label{sec:data}

\subsection{America On Line}
America On Line (AOL) is a company providing various types of Internet services (www.aol.com). Among them, AOL offers a search engine which allows to retrieve documents over the Web. We consider here a set of search queries performed on the AOL's search engine and officially released by the same company in $2006$~\footnote{The dataset is freely available at http://www.gregsadetsky.com/aol-data.}. The dataset consists of  $36\,389\,566$ queries performed by $657\,426$ different users over a period of three months (between $2006/03/01$ and $2006/05/31$). Several data are reported for each query: here we use only the identifier (ID) of the user performing the query and the time stamp indicating when the user performed the query (the resolution of the time stamps is in seconds).

\subsection{Ebay}

\begin{figure}
\includegraphics[width=0.5\textwidth]{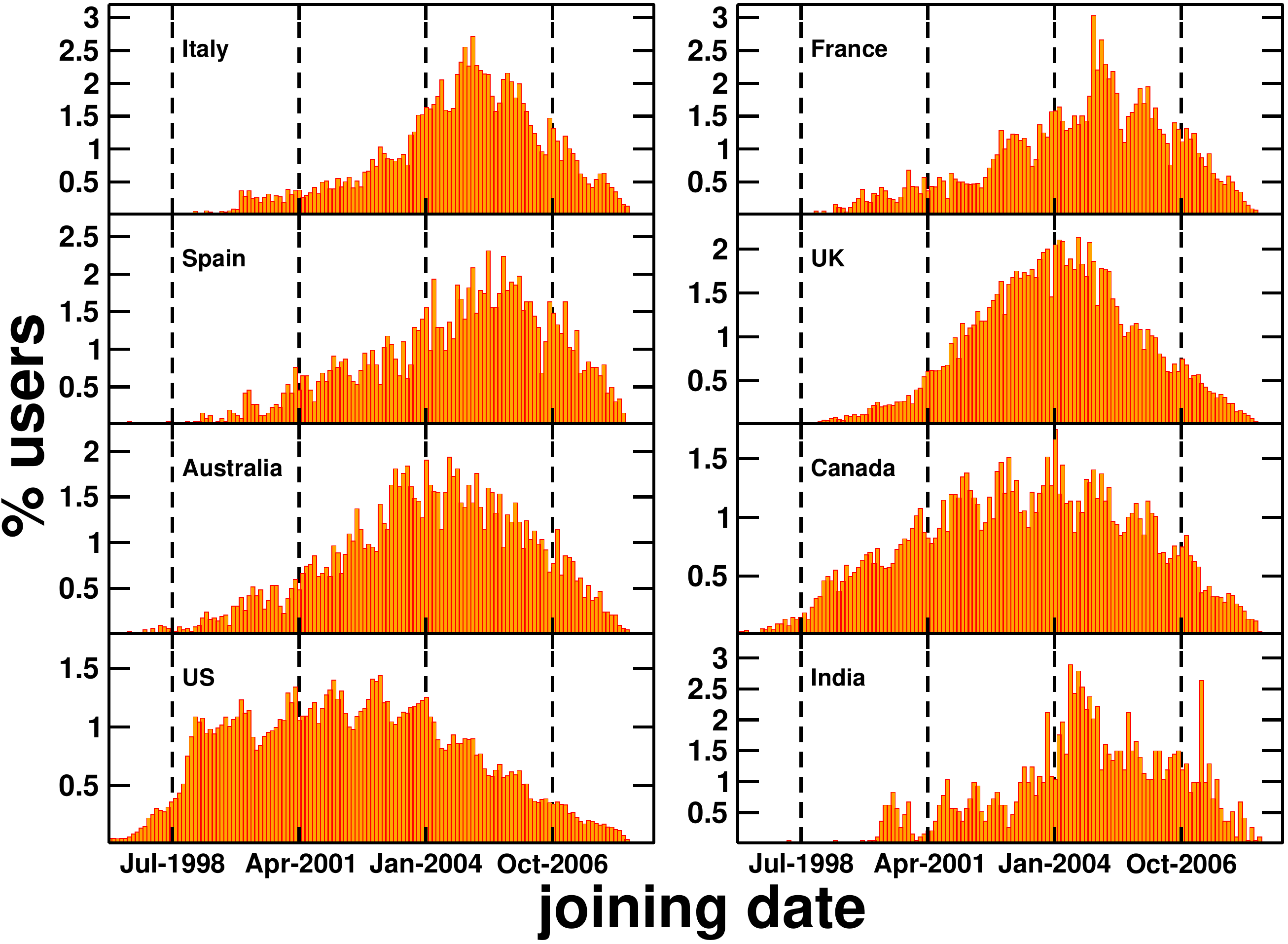}
\caption{(Color online) Percentage of users who have joined EB at a given date (time resolution is given in months). Each plot corresponds to a different country.}
\label{fig:join}
\end{figure}

Ebay (EB) is an on-line auction and shopping website in which people and businesses buy and sell goods and services worldwide (www.ebay.com). Born in $1995$ in the United States, EB has soon reached a great popularity and established localized websites in several other countries in the world. As an illustrative example, we plot in Fig.~\ref{fig:join} the percentage of users who have joined EB at a given time. This figure has only illustrative purposes since is representative only for a small portion of users~\footnote{It should be noticed that, despite its size, our  dataset represents a small portion of the whole population of Ebay, which is estimated to be  hundreds of millions large.}. The figure is however informative for the spreading of EB in the world: by following the peaks of registrations, we see that EB has first become popular in the US (peak in $2001$), then in English speaking countries (Australia, UK, Canada with peaks at the beginning of $2004$) and finally in the rest of the world (peaks in $2005$). 
\\ 
On EB, users sell or buy items via public auctions~\footnote{This is not always true since on-line shops sell goods without performing any auction.}. At the end of each auction, the user, who made the highest bid, pays the item and waits for receiving it. Sellers send items by using normal delivery services. After the buyer has received her/his good, she/he writes a feedback message about the transaction: she/he can decide to assign a positive, neutral or negative vote to the seller based on the quality of the object and the speed of the service. The seller can then reply with another feedback message which summarizes her/his opinion about the transaction. Feedback messages are made public through EB website and serve as quantitative measure for the reputation of buyers and sellers. The more positive feedback messages a user has received, the more reliable she/he is.
\\
We collected data directly from EB website~\footnote{The dataset can be found at http://filrad.homelinux.org.}. In order to download data with first selected four seed users and then followed the network of contacts (users are nodes of this network and feedback messages stand for directed connections between users), starting from our seeds up to their third shell. In this way, we downloaded $149\,087\,003$ feedback messages sent by $748\,282$ users. These data cover a period of more than ten years (from $1998$ to $2008$).  We stored data by using an anonymized ID for each user and the time stamp (with resolution in minutes) of each feedback message. For each user, we collected additional information as the country and the date of registration to EB (resolution in days), while for each feedback message we also registered the ID of the good correspondent to the transaction. It should be noticed that we consider only users which are not classified as ``shops'' or ``power sellers''. This roughly ensures the inclusion of only {\it normal} users with activity patterns typical of humans.

\subsection{Wikipedia}
Wikipedia (WP) is a free encyclopedia written in multiple languages and collaboratively created by volunteers. WP contains millions of articles and is currently the most popular general reference work on the Internet~\cite{alexa}. We consider the database containing all logging actions, performed by users, on the English website of WP (en.wikipedia.com~\footnote{Our dataset corresponds to the database dump of $2008/10/08$. Updated datasets are freely available at http://download.wikimedia.org.}). This dataset is composed of $17\,531\,208$ logging actions (i.e., uploads, deletes, etc.) performed by $7\,565\,401$ different users between $2004/12/23$ and $2008/10/08$.

\section{Activity statistics}
\label{sec:act}

\begin{figure*}[!hbt]
\includegraphics[width=0.3\textwidth, height=0.23\textwidth]{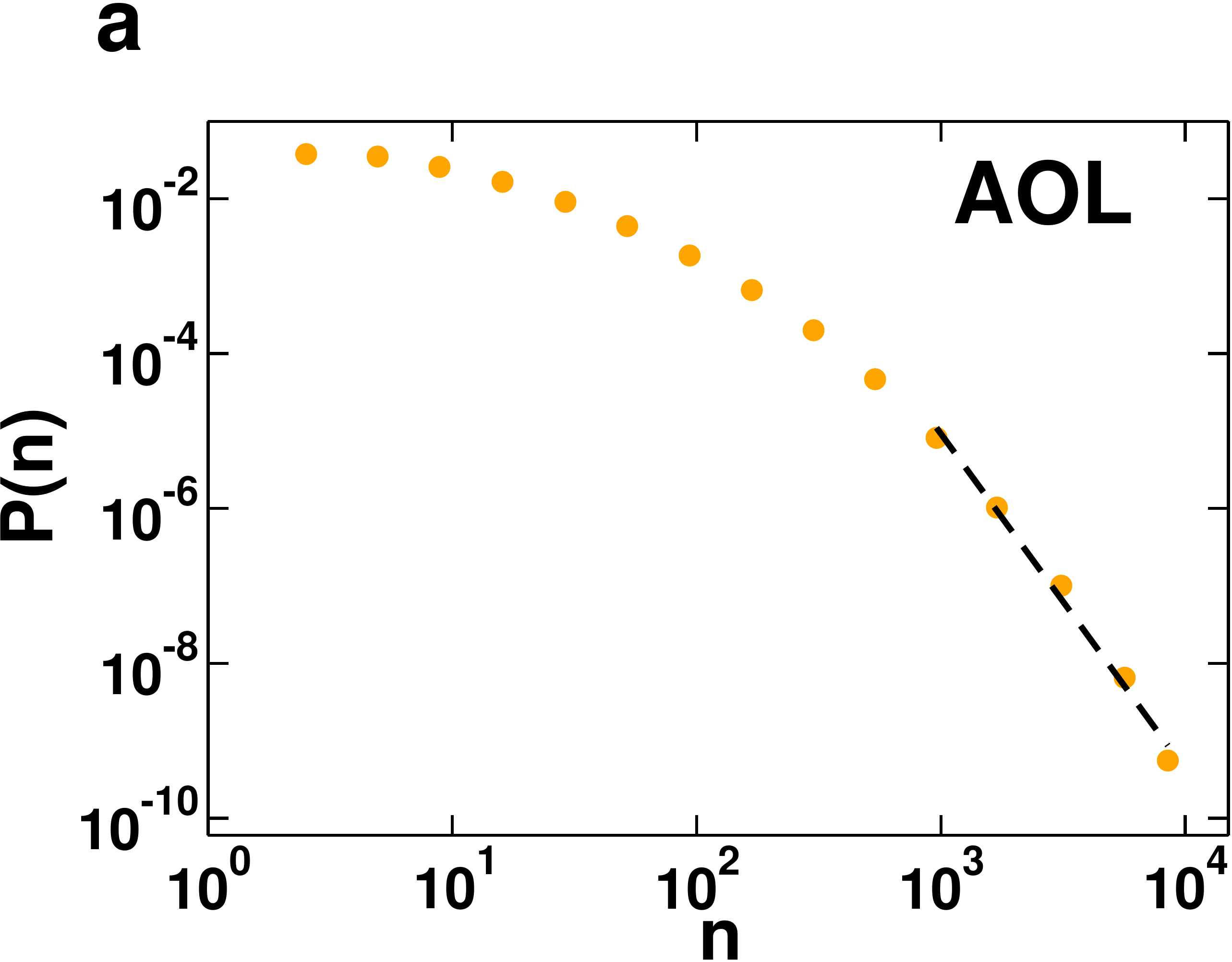}
\qquad
\includegraphics[width=0.3\textwidth, height=0.23\textwidth]{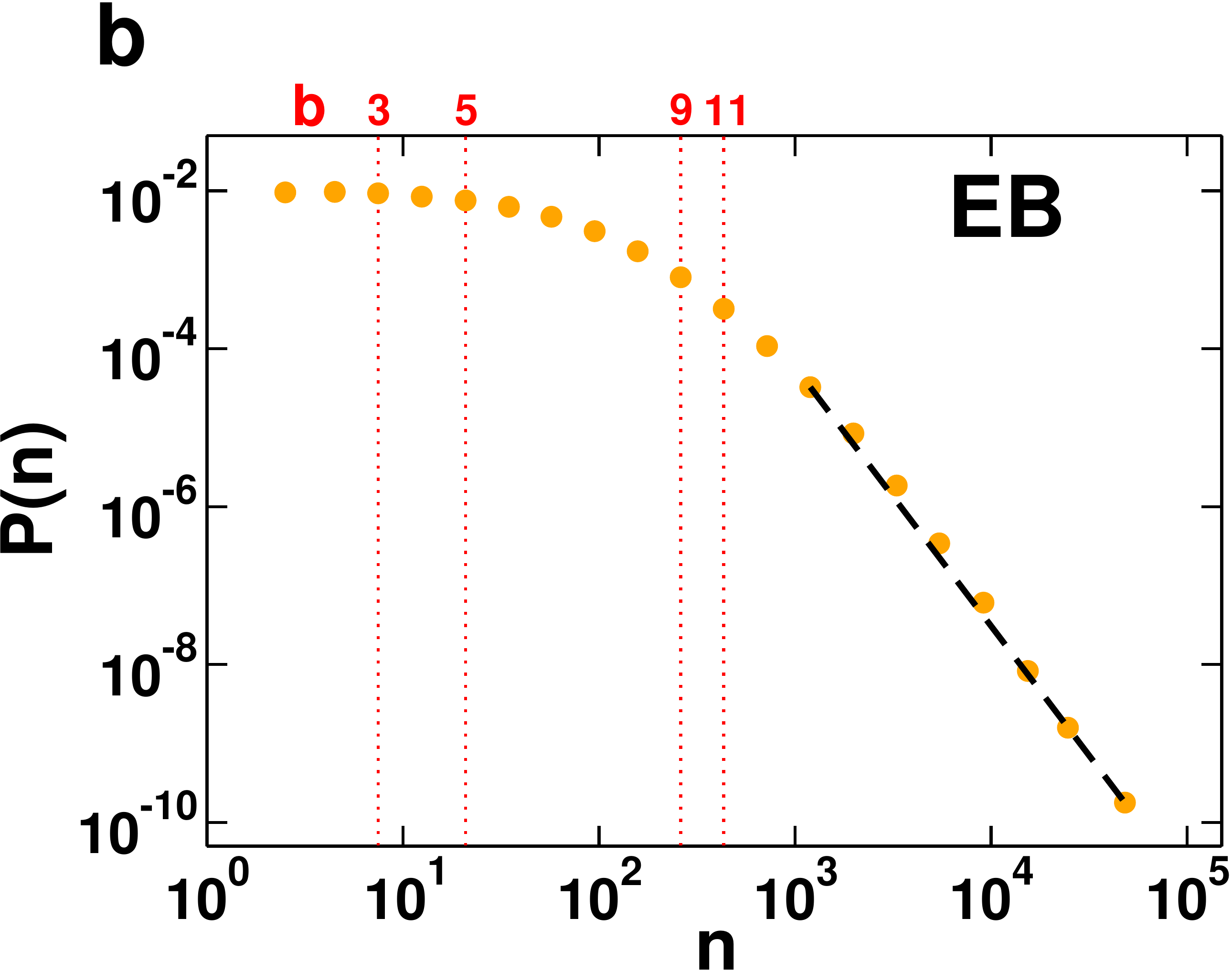}
\qquad
\includegraphics[width=0.3\textwidth, height=0.23\textwidth]{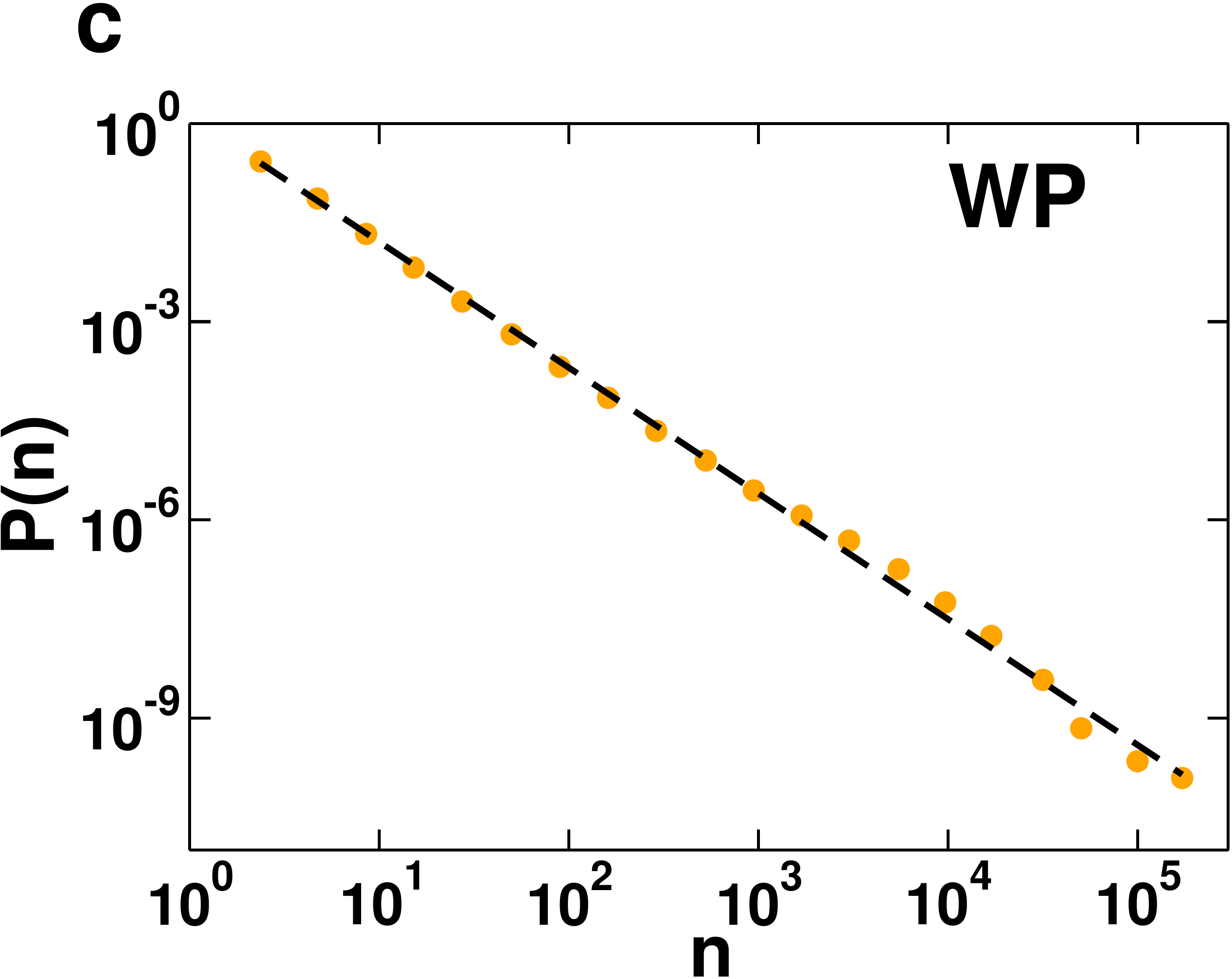}
\caption{(Color online) Fraction of users who have performed $n$ total number of operations [queries in (a), messages in (b) and logging actions in (c)]. In all cases, the tail of the distribution decays power-like as the total number of operations $n$ increases: $P\left(n\right) \sim 1/n^{\lambda}$ (dashed lines). The decay exponents are: $\lambda \simeq 4.3$ in (a), $\lambda \simeq 3.3$, (b) and $\lambda \simeq 1.9$ in (c). In all figures, points were obtained by using logarithmic binning. In (b) bins number $3, 5, 9$ and $11$ are evidenced since we will refer to them in Fig.s~\ref{figks2}a and~\ref{fig:scale}.}
\label{fig1}
\end{figure*}

In Fig.s~\ref{fig1} we plot  the probability $P\left(n\right)$, calculated as the relative (with respect to the whole population) number of users who have performed $n$ total operations. For all databases analyzed in this paper, we see that $P\left(n\right)$ is broad and its tail decays power-like as the $n$ increases [i.e., $P\left( n \right) \sim n^{-\lambda}$, for $n \gg 1$]. The decay exponents are: $\lambda \simeq 4.3$ for AOL, $\lambda \simeq 3.3$ for EB and $\lambda \simeq 1.9$ for WP. In the case of AOL the value of the exponent suggests a decay which is more exponential than power-like (see Fig.~\ref{fig1}a), differently in the case of the WP's dataset,  $P\left(n \right)$ fits very well a power-law function for every value of $n$ and not just along the tail (see Fig.~\ref{fig1}c).
\\
These results tell us that users, involved in Web activities, are heterogeneous since the number of operations $n$ (queries, messages or logging actions, depending on the dataset) widely changes among them. This fact is particularly relevant because, as we will see in the rest of the paper, the number of operations performed by a user plays an important role for the determination of her/his activity pattern.

\section{Inter-event time statistics}
\label{sec:time}

\subsection{Global inter-event time distribution}
\label{sec:time_a}

\begin{figure*}[!hbt]
\includegraphics[width=0.3\textwidth, height=0.23\textwidth]{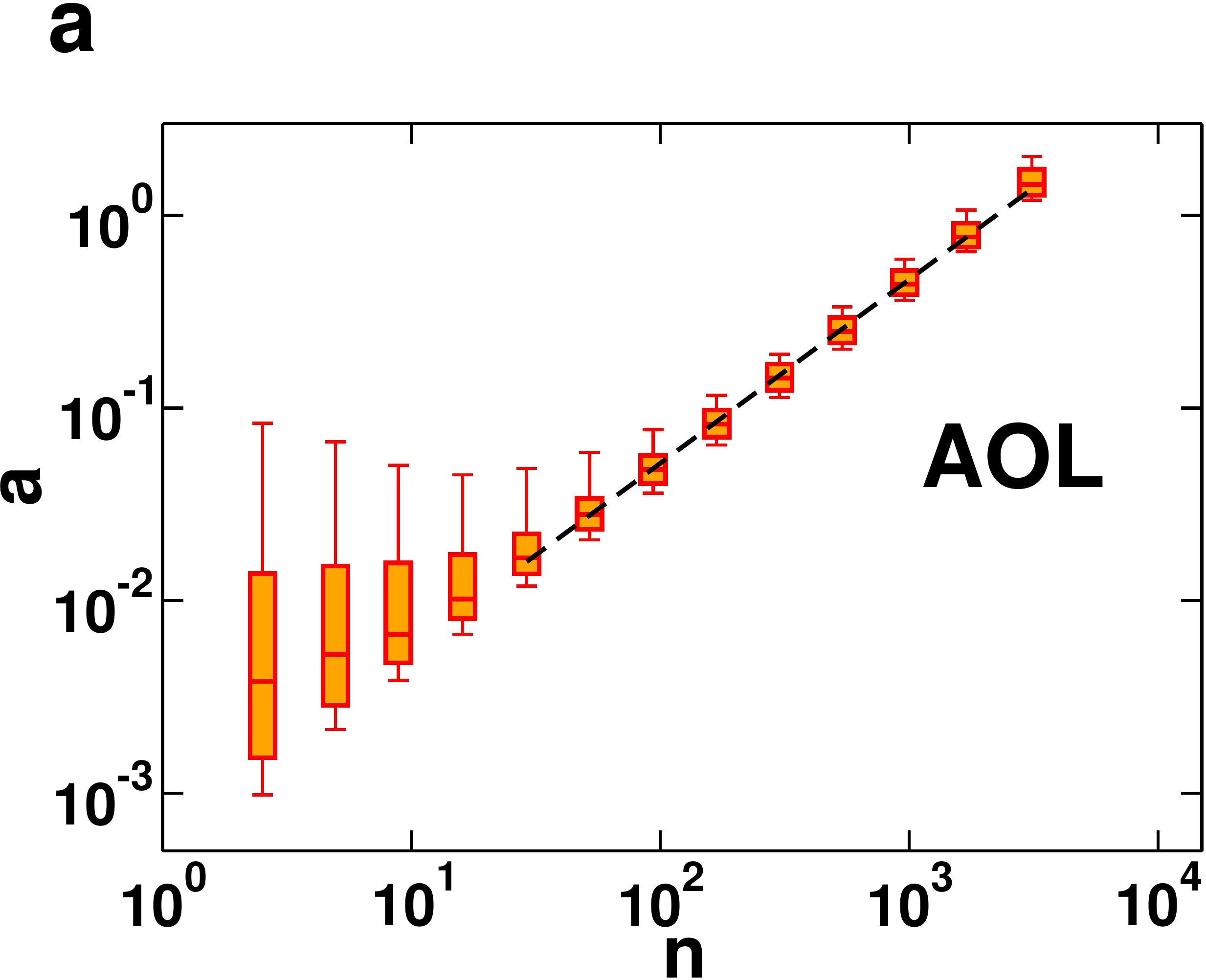}
\qquad
\includegraphics[width=0.3\textwidth, height=0.23\textwidth]{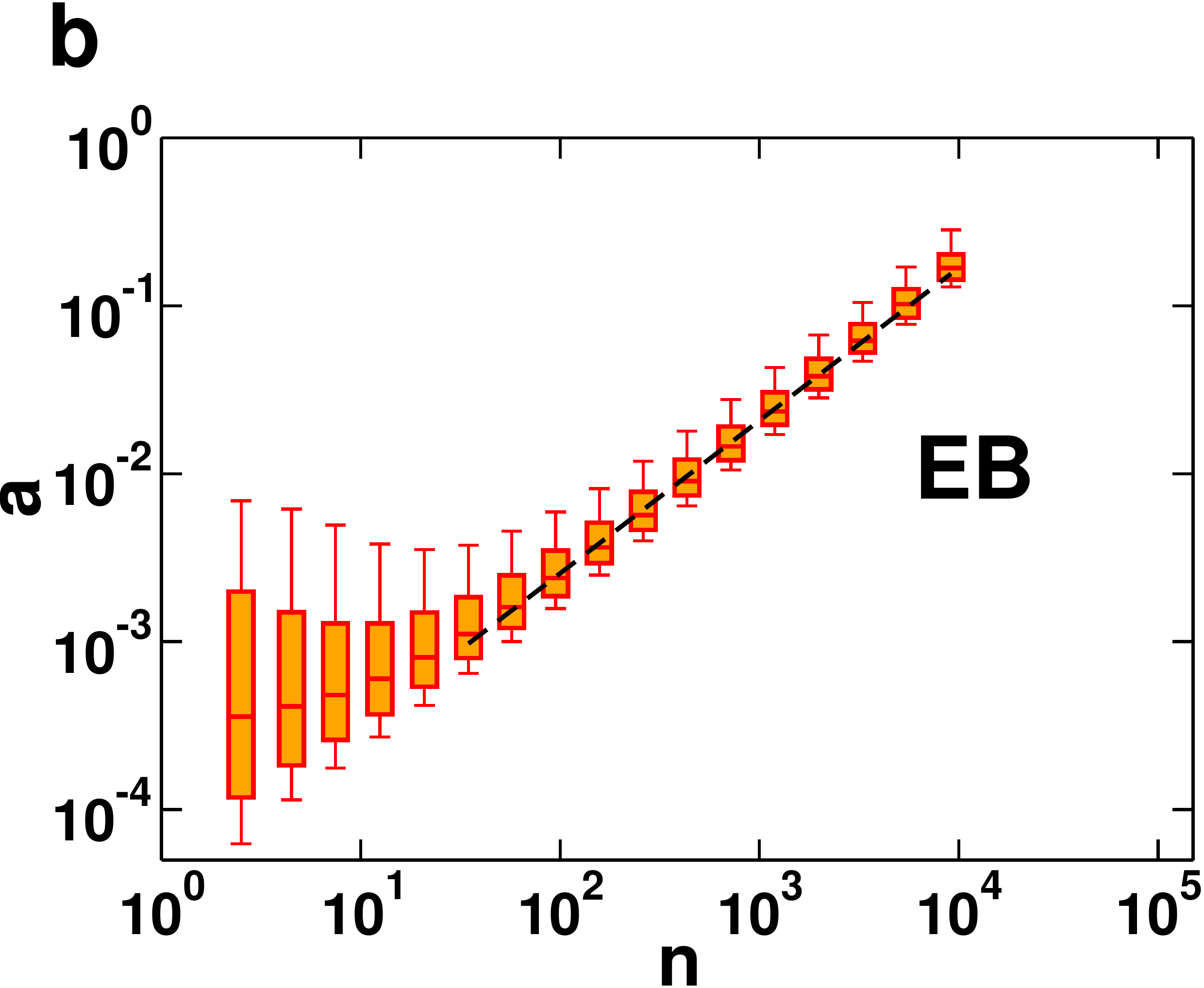}
\qquad
\includegraphics[width=0.3\textwidth, height=0.23\textwidth]{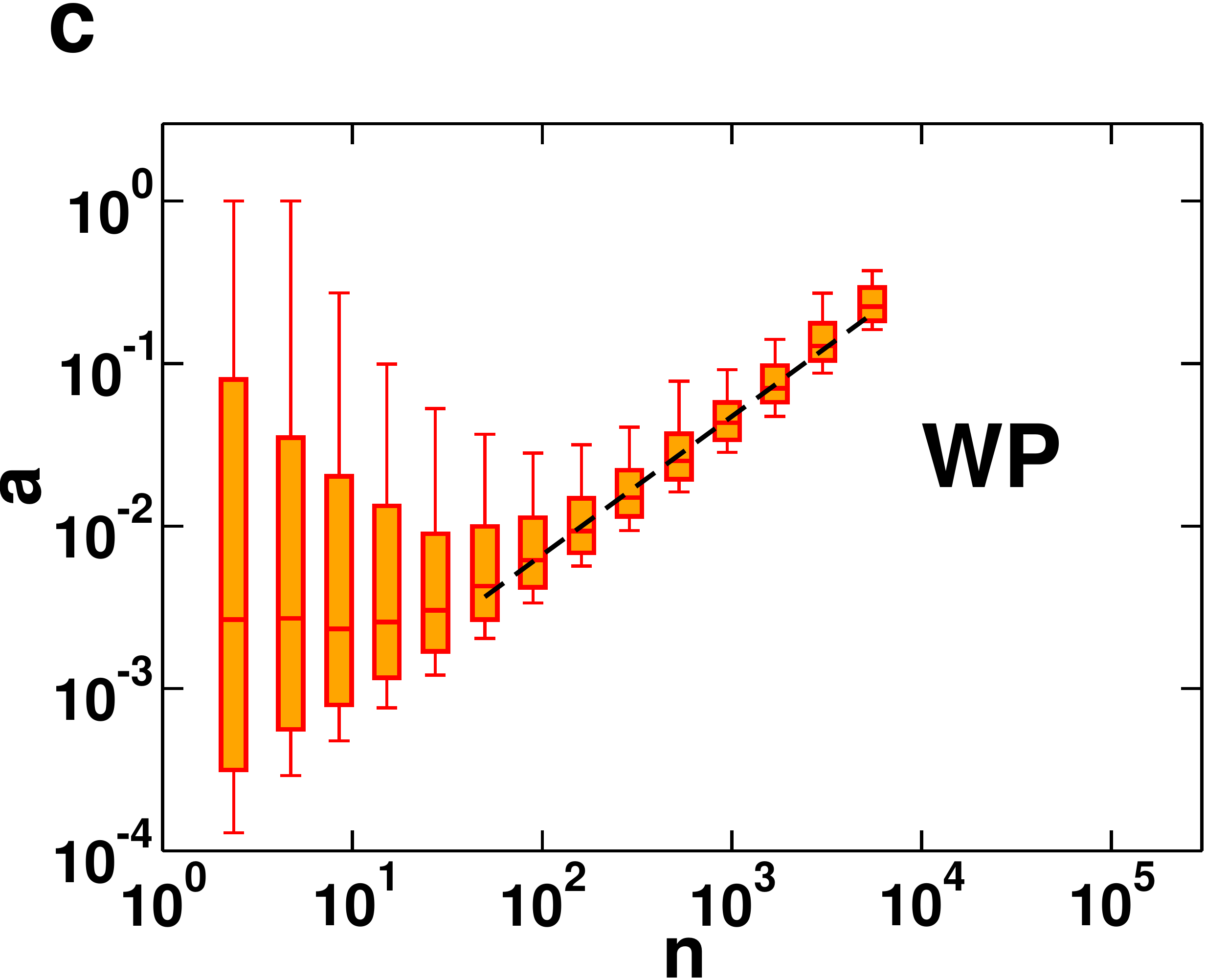}
\caption{(Color online) Activity $a$ as a function of the number of operations $n$ performed by users. Activities are expressed in number of operations per hour. In all plots, users have been grouped into bins and the values of $a$ corresponding to the top $50\%$ (horizontal bars), $25\%$ and $75\%$ (boxes) and $10\%$ and $90\%$ (error bars) of the population are shown for each bin. For large values of $n$, $a$ grows almost linearly with $n$ (dashed lines have slope close to one). The bin divisions are the same of those reported in Fig.s~\ref{fig1}. Only bins populated by at least $100$ users are considered and shown in these plots.}
\label{fig2}
\end{figure*}

Suppose the user $i$ has performed $n_i$ operations at the instants of  time $t_{i_1}, t_{i_2}, t_{i_3}, \ldots, t_{i_{n_i}}$, where $t_{i_1}\leq t_{i_2} \leq t_{i_3} \leq \ldots t_{i_{n_i}}$. This information allows to compute the inter-event time between subsequent operations: $\tau_{i_1}=t_{i_2}-t_{i_1}, \tau_{i_2}=t_{i_3}-t_{i_2}, \ldots,  \tau_{i_{n_i-1}}=t_{i_{n_i}}-t_{i_{n_i-1}}$. In general, the interval of time $\tau$ between two subsequent operations strongly depends on how much the considered user is active.
\\
Users performing a large number of operations are very active, in the sense that the average time gap between two subsequent operations is small. In order to quantify this observation, we define  the average activity $a_i$ of the user $i$ as
\begin{equation}
a_i = \frac{n_i}{t_{i_{n_i}} - t_{i_1}}\;\;\;,
\label{eq:activity}
\end{equation}
where  $n_i$ is the total number of operations performed by the user $i$ and $t_{i_{n_i}} - t_{i_1}$ is the length of the interval of time in which the user $i$ is active. We consider only users who have performed at least two actions in a period of activity larger than one hour (i.e., all users $i$ satisfying $n_i \geq 2$ and  $t_{i_{n_i}} - t_{i_1} \geq 1 \;\textrm{hour}$.). This restricts the calculations to $557\,513$ users in AOL and  $733\,335$ and $292\,799$ users in EB and in WP, respectively. Fig.s~\ref{fig2} show the relation between the average activity and the number of operations. Data have been grouped into equally spaced, on the logarithmic scale, bins. We compute the values of $a$ corresponding to the top $10\%$, $25\%$, $50\%$, $75\%$ and $90\%$ of the population of each bin. Only bins populated by at least $100$ users are shown. For small values of $n$, $a$ has large fluctuations, while fluctuations become smaller as $n$ increases. In general, $a$ and $n$ are linearly correlated. It should be noticed that $a_i$ is equivalent to the inverse of the average inter-event time since $t_{i_{n_i}} - t_{i_1} = \sum_{q=1}^{n_i-1} \tau_{i_q}$.

\

The probability $P_i\left(\tau\right)$, that two subsequent operations performed by the  $i$-th user differ by $\tau$ units of time, can be calculated as  
\begin{equation}
P_i\left(\tau\right) = \frac{1}{n_i-1} \sum_{q=1}^{n_i-1} \delta_{\tau, \tau_{i_q}} = \frac{x_i\left(\tau\right)}{n_i-1}\;\;\;,
\label{eq:tau_user}
\end{equation}
where $\delta_{r,s}$ is the Kronecker delta which equals one if $r=s$ and zero otherwise. $x_i\left(\tau\right)$ stands for the total number of subsequent operations, which differ by $\tau$, performed by the user $i$. The normalization of eq.(\ref{eq:tau_user}) is preserved since $\sum_\tau x_i\left(\tau\right) = n_i-1$.
\\
If the population is composed of $N$ users, the probability $P\left(\tau\right)$ that a generic user performs  two subsequent operations which differ by an amount of time $\tau$ is given by
\begin{equation}
P\left(\tau\right) = \frac{\sum_{i=1}^{N} x_i\left(\tau\right)}{\sum_{\eta=0}^{\tau_M} \sum_{i=1}^{N} x_i\left(\eta\right)} = \frac{\sum_{i=1}^{N} \left(n_i -1 \right)P_i\left(\tau\right)}{\sum_{i=1}^{N}  \left(n_i -1 \right)} \;\; ,
\label{eq:tau_population}
\end{equation}
where $\tau_M$ is the maximal value of $\tau$ observed in the dataset. It is important to notice that eq.(\ref{eq:tau_population}) represents the best estimate for the inter-event time probability distribution function (pdf) in the hypothesis that all $P_i\left(\tau\right)$ are the same and basically corresponds to the weighted average of the single  users'pdfs.
\\

\begin{figure*}[!hbt]
\includegraphics[width=0.3\textwidth, height=0.23\textwidth]{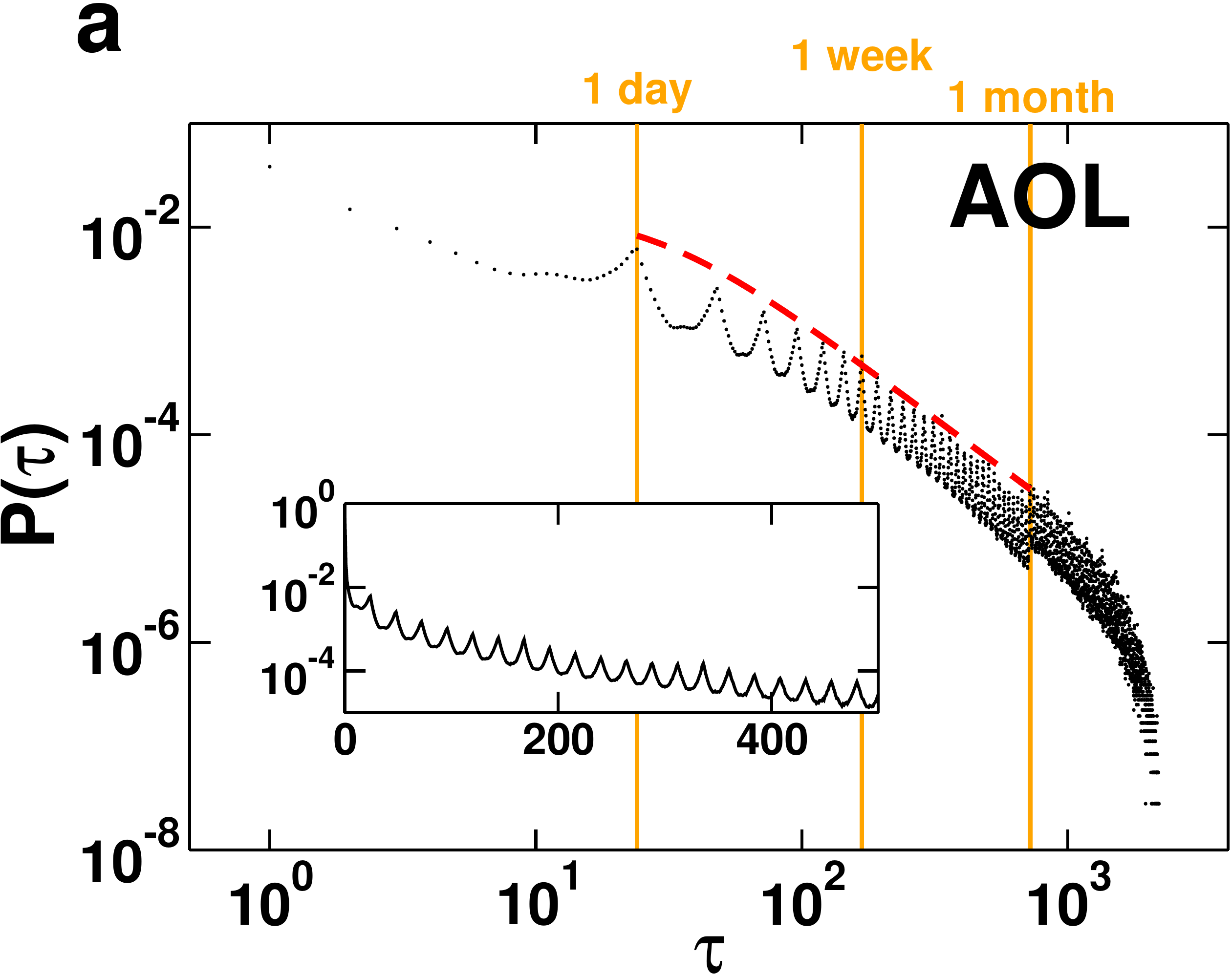}
\qquad
\includegraphics[width=0.3\textwidth, height=0.23\textwidth]{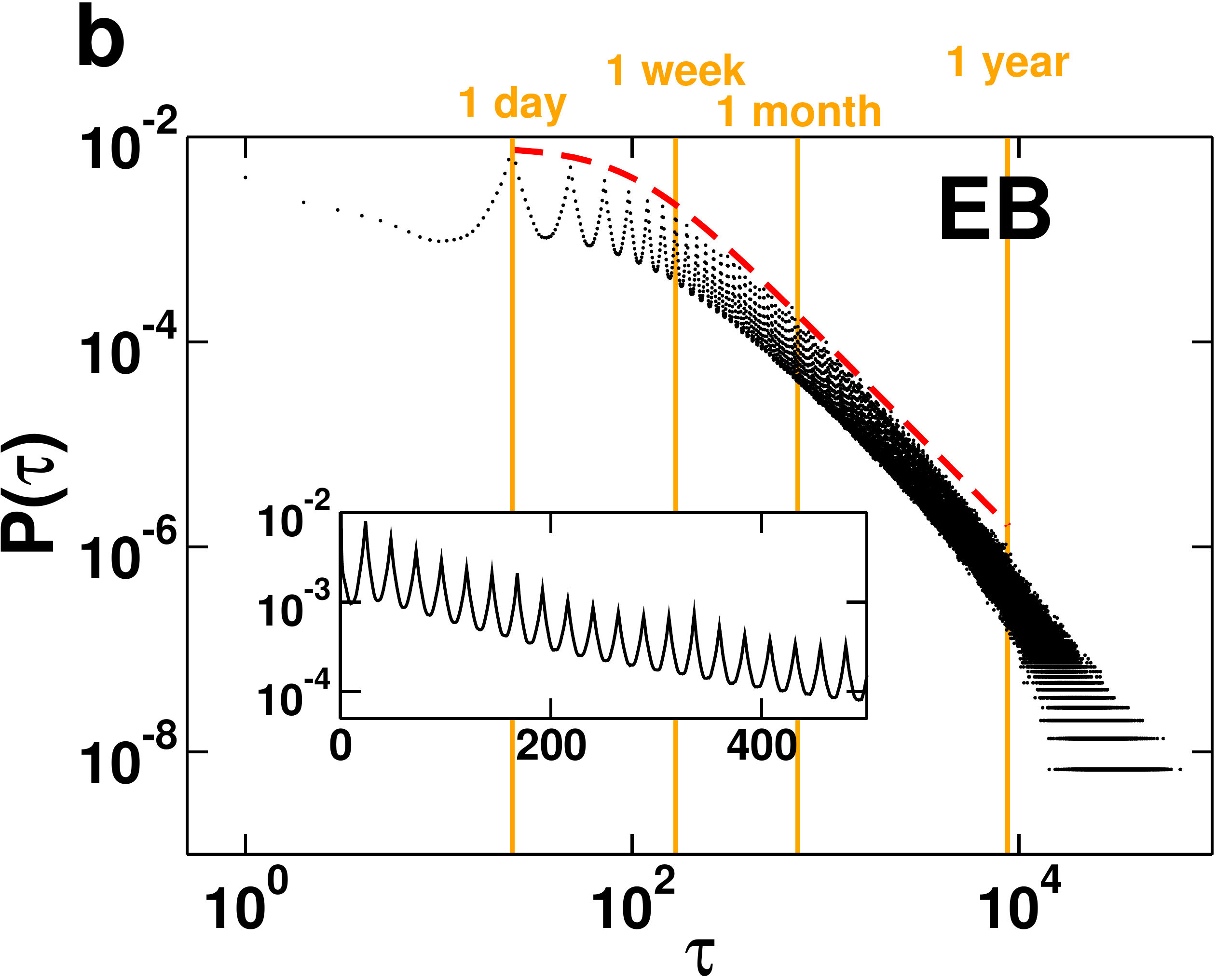}
\qquad
\includegraphics[width=0.3\textwidth, height=0.23\textwidth]{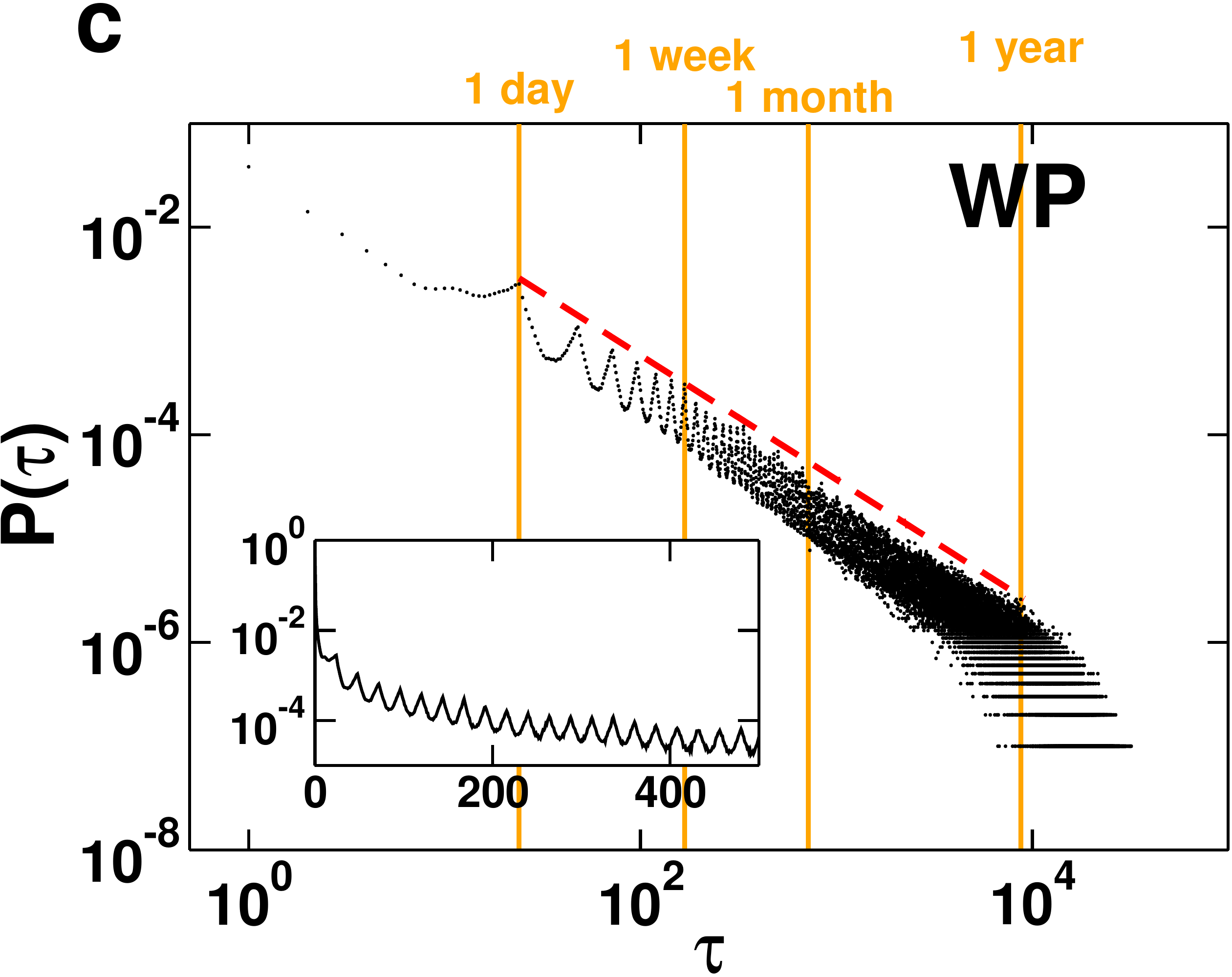}
\caption{(Color online) The main plots show the probability $P\left(\tau\right)$ that a user performs two subsequent operations [queries in (a), messages in (b) and logging actions in (c)] at time difference equal to $\tau$. $P\left(\tau\right)$ is averaged over all users by using eq.(\ref{eq:tau_population}). In all cases, $P\left(\tau\right)$ decays power-like as described by eq.(\ref{eq:empirical}), and the decay exponents (dashed lines) are: $\beta \simeq 1.9$ in (a), $\beta \simeq 1.9$ in (b) and $\beta \simeq 1.2$ in (c). The insets show a zoom of $P\left(\tau\right)$ from which it is possible to clearly observe periodic (daily and weekly) oscillations.}
\label{fig3}
\end{figure*}

The global pdfs $P\left(\tau\right)$ calculated for  AOL, EB and WP are reported in the main plots of Fig.s~\ref{fig3}. In order to have much cleaner figures, we express $\tau$ with a resolution of hours. It should  be noticed that in all cases the most probable value is $\tau=0$, since $P\left(0\right)>P\left(\tau\right) \;\;,\; \forall \; \tau>0$, which means that the majority of subsequent operations has time difference smaller than thirty minutes. In particular we have: $P\left(0\right) \simeq 0.71$ for AOL, $P\left(0\right) \simeq 0.58$ for EB and $P\left(0\right) \simeq 0.77$ for WP. As we can clearly see from Fig.s~\ref{fig3}, the global inter-event time pdfs present a power-law decay 
\begin{equation}
P\left( \tau\right) \sim a/\left[1 + \left(\tau/b\right)^\beta \right] \;\;\; ,
\label{eq:empirical}
\end{equation}
with decay exponents equal to $\beta \simeq 1.9$ for AOL, $\beta \simeq 1.9$ for EB and $\beta \simeq 1.2$ for WP.

\subsection{Reliability of $P\left( \tau \right)$}
\label{sec:time_b}

\begin{figure*}
\includegraphics[width=0.3\textwidth, height=0.23\textwidth]{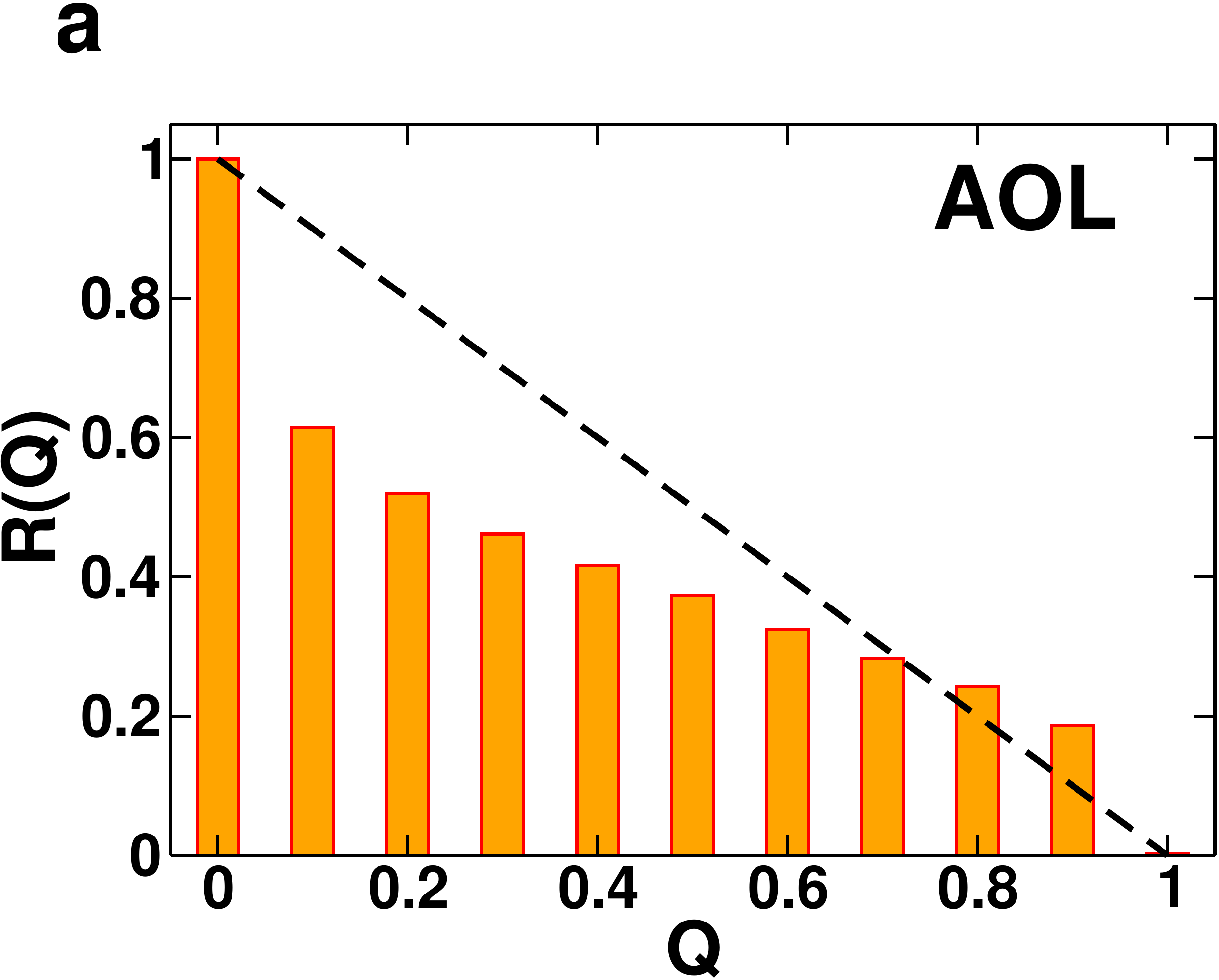}
\qquad
\includegraphics[width=0.3\textwidth, height=0.23\textwidth]{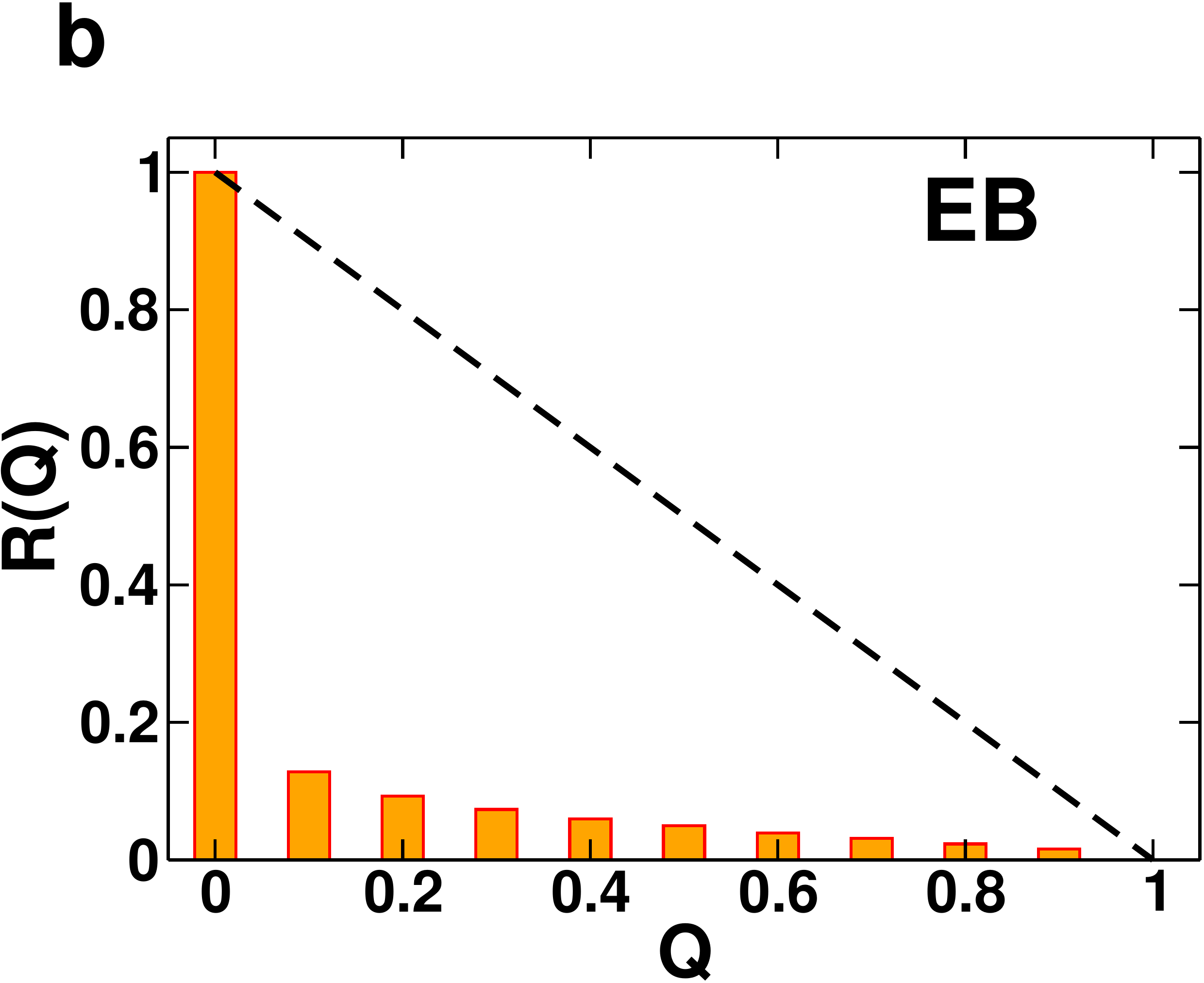}
\qquad
\includegraphics[width=0.3\textwidth, height=0.23\textwidth]{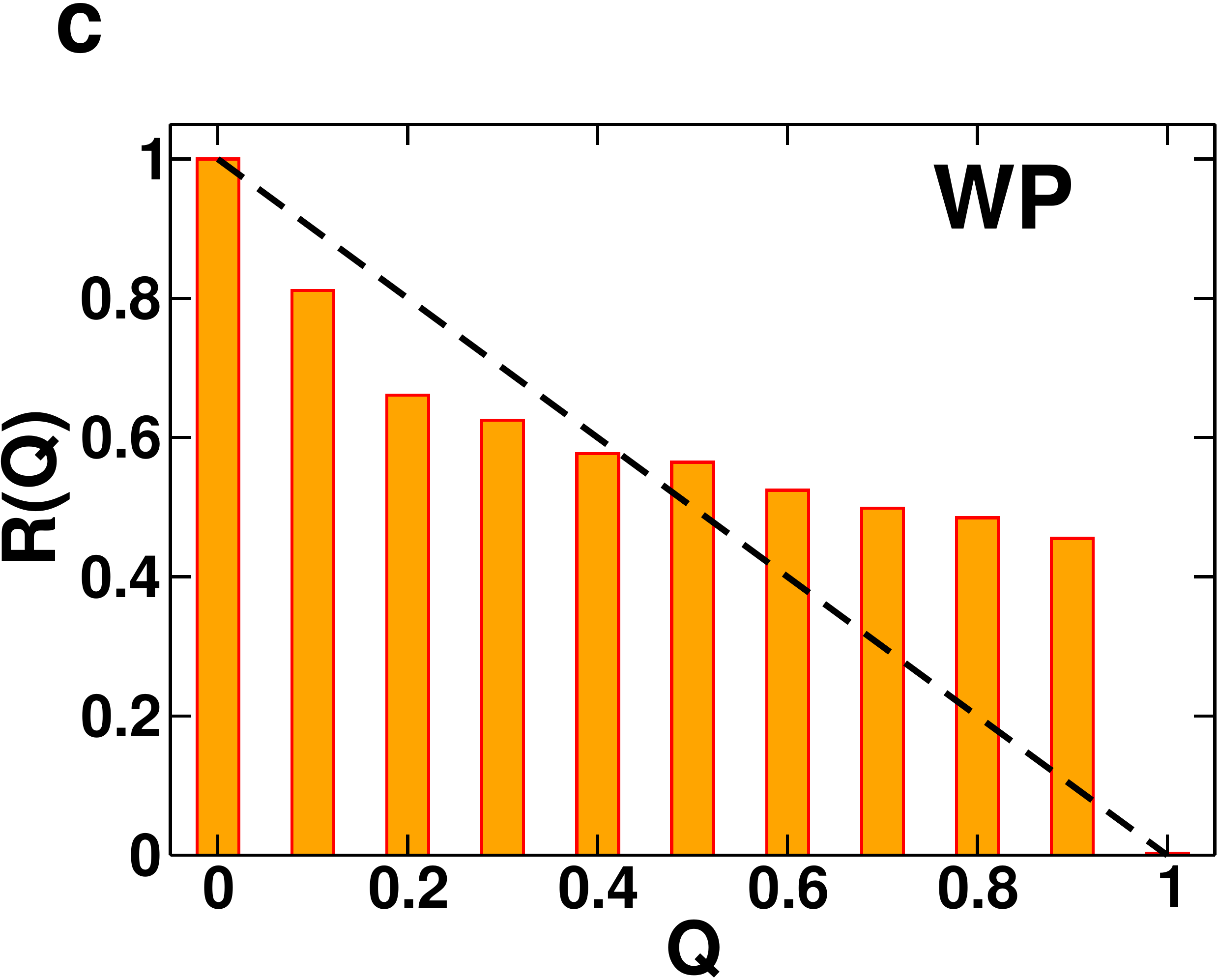}
\caption{(Color online) We report the fraction of users $R\left(Q\right)$ whose inter-event time pdf is described by the the global $P\left( \tau \right)$ with significance level larger or equal to $Q$. In all figures, dashed lines stand for the function $1-Q$, which is the expected behavior of $R\left(Q\right)$.}
\label{figks}
\end{figure*}

\begin{figure*}[!hbt]
\includegraphics[width=0.3\textwidth, height=0.23\textwidth]{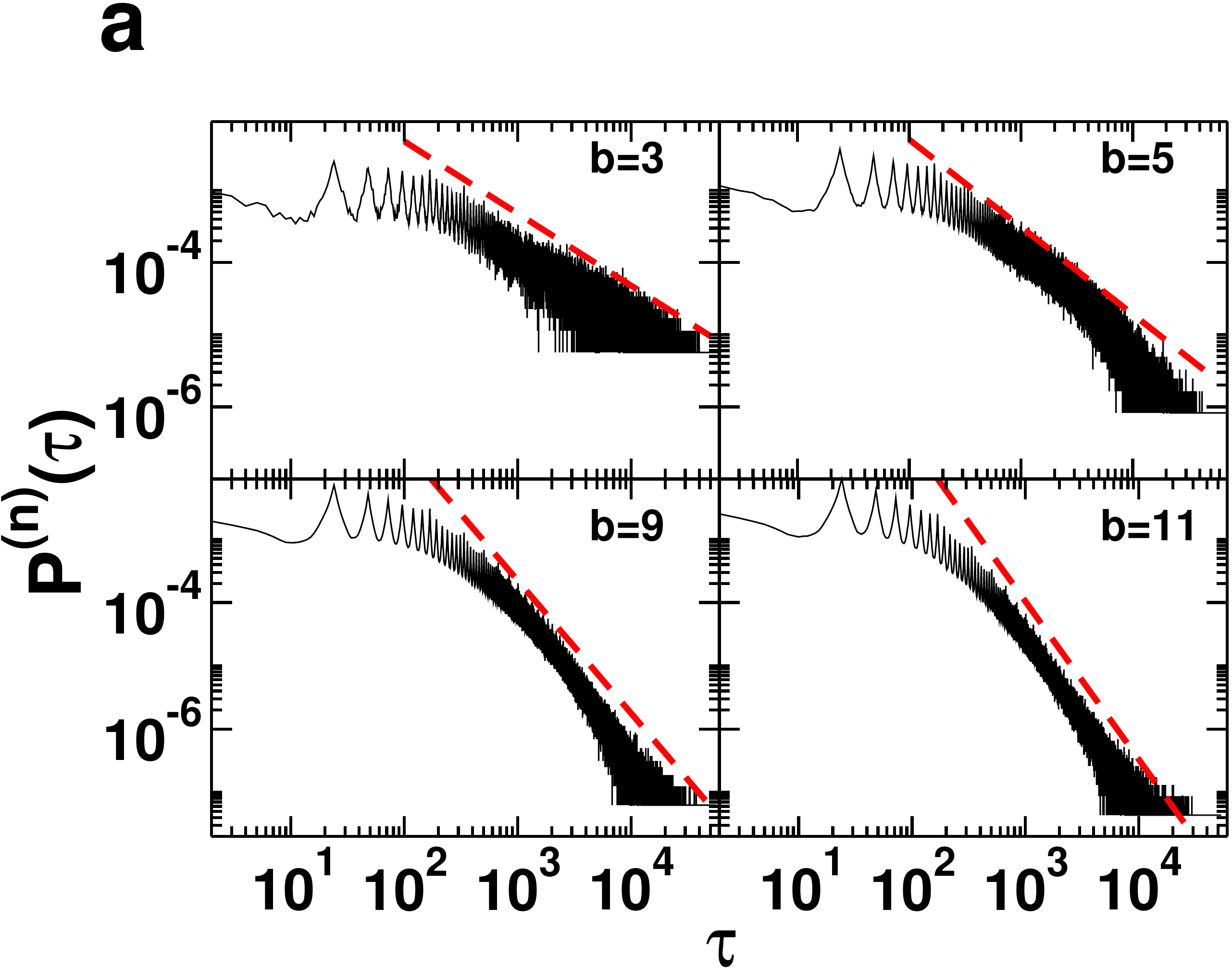}
\qquad
\includegraphics[width=0.3\textwidth, height=0.23\textwidth]{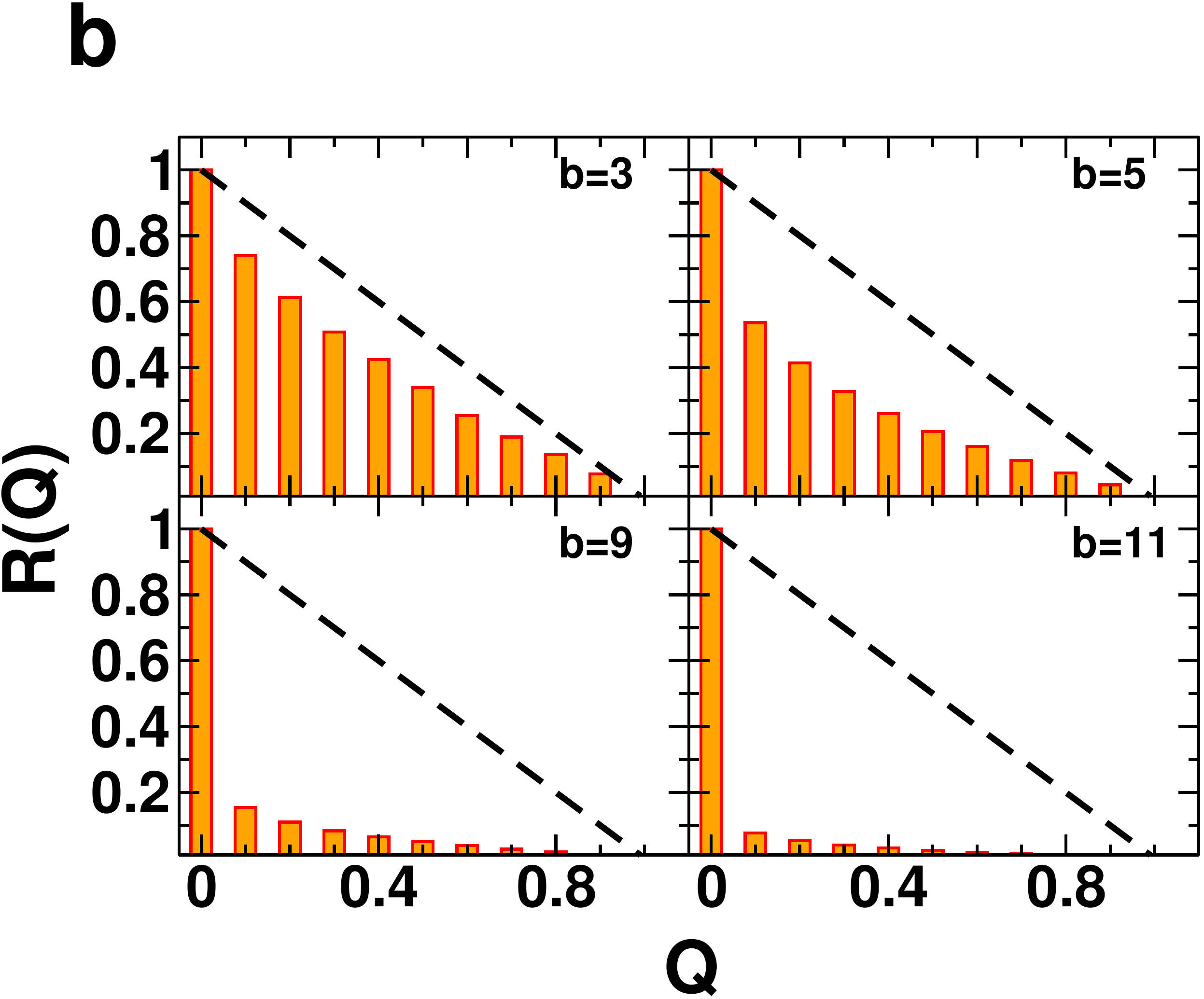}
\qquad
\includegraphics[width=0.3\textwidth, height=0.23\textwidth]{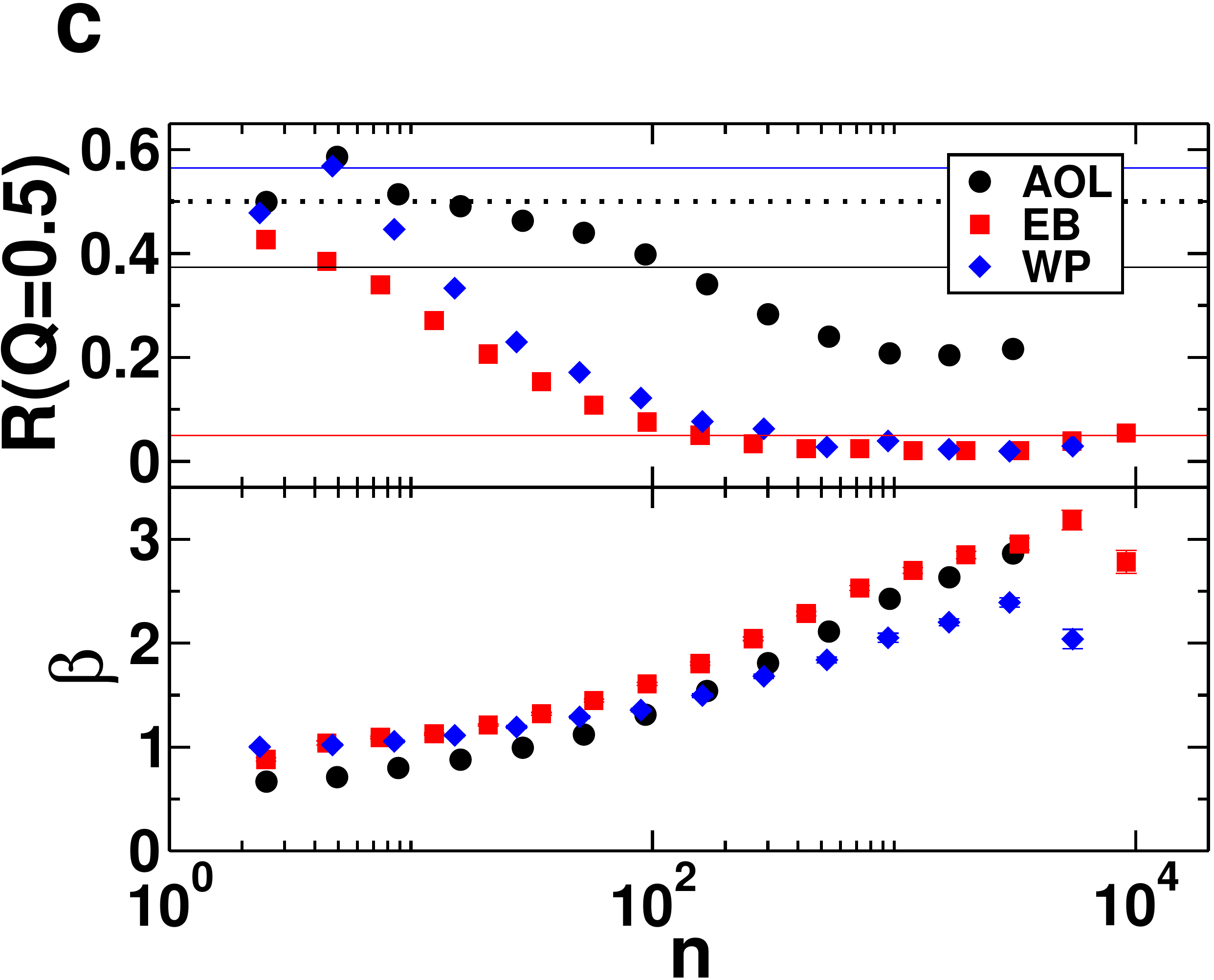}
\caption{(Color online) (a) Inter-event time pdfs $P^{\left(n\right)}\left(\tau \right)$ for users with the same total number of operations. Each panel corresponds to a set of users who have performed a similar number of operations. We consider the EB datasets and plot the $P^{\left(n\right)}\left(\tau \right)$ corresponding to the bins $b=3, 5, 9$ and $11$ of Fig.~\ref{fig1}b. Dashed lines stand for best fit power-laws with decay exponents $\beta = 1.1, 1.2, 1.8$ and $2.3$ for the cases $b=3, 5, 9$ and $11$, respectively. (b) In each panel we report the fraction $R\left(Q\right)$ of users whose inter-event time pdf is described by $P^{\left(n\right)} \left(\tau \right)$ with a probability at least equal to $Q$. We consider the same bins as those analyzed in (a). (c) In the top panel, $R\left(Q=0.5\right)$ is plotted as a function of $n$. Measured values are plotted as black circles for AOL, red squares for EB and blue dyamonds for WP. Horizontal lines stand for comparison: the dotted line is the expected value of $R\left(Q=0.5\right)$, solid lines are the values of $R\left(Q=0.5\right)$ calculated in the case of the global pdfs, from top to bottom WP (blue), AOL (black) and EB (red), respectively (see Fig.s~\ref{figks}). The degree of compatibility  between inter-event time pdfs of users who have performed a similar number of operations decreases as $n$ increases. In the bottom panel, we report the value of the decay exponent $\beta$ for $P^{\left(n\right)} \left(\tau \right)$ as a function of $n$. It is interesting to notice that $\beta$ follows almost the same behavior in all databases.}
\label{figks2}
\end{figure*}

$P\left(\tau\right)$ has been calculated as the weighted average of the inter-event time pdfs of  single users. As already stated, eq.(\ref{eq:tau_population}) is the most representative way to calculate $P\left(\tau\right)$ only in the hypothesis that all users behave in a similar way.
\\
In order to test the reliability of $P\left(\tau\right)$ as probability for the inter-event time statistics of each user we make use of the Kolmogorov-Smirnov (KS) test~\cite{mood74}. KS is non-parametric statistical test which allows to quantify to which extent the hypothesis that two pdfs were drawn from the same underlying distribution is valid. In our specific case, we calculate for each user the cumulative distribution function (cdf) $C_i \left( \tau \right) = \sum_{\eta=0}^{\tau} \; P_i \left( \eta \right)$ and we perform a KS test, comparing this cdf with the one valid for the whole population  $C \left( \tau \right) = \sum_{\eta=0}^{\tau} \; P \left( \eta \right)$. From the KS test we obtain a number $0 \leq Q \leq 1$ which basically quantifies the significance level of similarity between the two distributions: high values of $Q$ mean that is very probable that the two sets of data have been generated from the same underlying distribution, differently a small $Q$ tells that the hypothesis of having a common underlying distribution is unlikely.
\\
As we can see from Fig.s~\ref{figks}, in general $P\left(\tau\right)$ does not well represent the activity of single users. In these figures, we consider the quantity $R\left(Q\right)$, which stands for the normalized number of users whose inter-event time pdf is described by $P\left(\tau\right)$ with a significance level larger or equal to $Q$. Since  $R\left(Q\right)$ is the complementary cdf of the KS cdf, we expect that $R\left(Q\right) = 1-Q$. From Fig.s~\ref{figks}, we see obviously that $R\left(Q\right)$ is a decreasing function of $Q$, but that it does not follow the expected behavior. It should be noticed that, in the case of WP, $R\left(Q\right)$ follow a functional form very similar to the expected one, but this may be an artifact due to the shape of the correspondent $P(n)$: the global inter-event time pdf is mainly due to the contribution of users with small $n$ and the same poorly active users are those who contribute mainly to the value of $R\left(Q\right)$. Just to a give a quantitative idea, we can for example say that the percentage of users whose inter-event time statistics is described by $P\left(\tau\right)$ with a significance of $50\%$ are $37\%$ for AOL, $5\%$ for EB and $56\%$ for WP while from KS statistics we expect to have $50\%$.

\

The main problem is that the inter-event time pdf of a user is strictly dependent on  the total number of operations performed by the same user~\cite{zhou08} and the pdf of the number of actions performed is wide (see Fig.s~\ref{fig1}). We therefore consider the inter-event time pdf $P^{\left(n\right)}\left(\tau \right)$ of users with the same number of operations $n$. For simplicity, we divide the entire population in $20$ sets of users with similar total number of operations. The divisions corresponds exactly to those used in Fig.s~\ref{fig1}, where users are placed into equally spaced bins on the logarithmic scale depending on the total number of actions $n$ they have performed~\footnote{The results are in general quantitatively dependent on the number of bins and the way in which the bins are determined. However, qualitatively analogous results can be obtained for other choices of the bins.}. We then compute the $P^{\left(n\right)}\left(\tau \right)$ corresponding to each of these bins. In Fig.~\ref{figks2}a, we consider the EB dataset and plot the inter-event time pdfs corresponding to four different bins: $b=3, 5, 9$ and $11$ (which correspond to average numbers of operations equal to $\langle n \rangle = 7.5, 20.8, 157.4$ and $432.9$, respectively). As one can see, $P^{\left(n\right)}\left(\tau \right) \sim \tau^{-\beta}$ in all cases, but the decay exponent $\beta$ changes as a function of $n$: in the represented cases, we have for example $\beta \simeq 1.1, 1.2, 1.8$ and $2.3$, respectively. In general, $P^{\left(n\right)}\left(\tau \right)$ well describes the statistics associated with the inter-event times of single users with $n$ total actions (Fig.~\ref{figks2}b). We calculate the quantity $R\left(Q\right)$ also in this case and we find that the percentages of users whose inter-event time pdf is described by $P^{\left(n\right)}\left(\tau \right)$ with a significance larger than $Q=0.5$ are: $34\%, 21\%, 5\%$ and $3\%$. In general, users with a reasonable small total number of operations behave similarly and $P^{\left(n\right)}\left(\tau \right)$ well represents the statistics associated with their activity. Differently, for large values of $n$, each user behaves in her/his own way and the statistics of her/his inter-event times differ from those of the other users with the same number of operations. The same qualitative results are valid also for AOL and WP. Fig.~\ref{figks2}c summarizes our analysis. In the top panel, the ratio $R\left( Q=0.5\right)$ of users whose inter-event time pdf is described by  $P^{\left(n\right)}\left(\tau \right)$ with an accuracy  larger that $Q=0.5$ is plotted as a function of $n$. In the bottom panel, the decay exponent $\beta$ is plotted as a function of $n$. In general we see that $R$ decreases while $\beta$ becomes larger as $n$ increases.

\subsection{Scaling of inter-event probability distributions}
\label{sec:scaling}

\begin{figure}
\includegraphics[width=0.45\textwidth, height=0.345\textwidth]{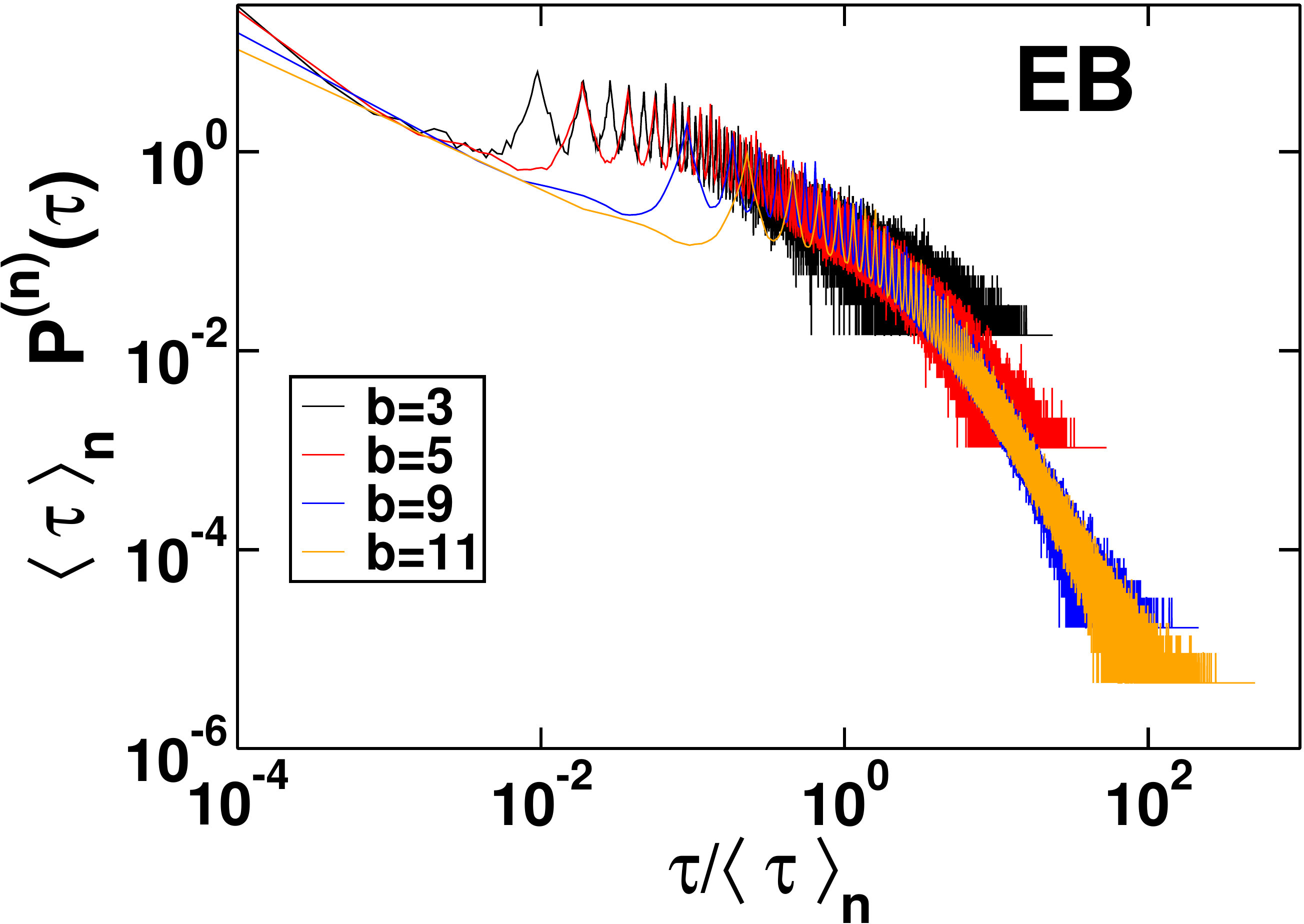}
\\
\caption{(Color online) (a) Scaling of the inter-event time distributions $P^{(n)}\left(\tau\right)$ in the case of the EB dataset. Data are the same as those already plotted in Fig.~\ref{figks2}a, but now each pdf $P^{(n)}\left(\tau\right)$ is appropriately rescaled with the average inter-event time $\langle \tau \rangle_n$ of the respective population of users. The scaling produces a nice collapse between the different curves.
}
\label{fig:scale}
\end{figure}

\begin{figure*}
\includegraphics[width=0.3\textwidth, height=0.23\textwidth]{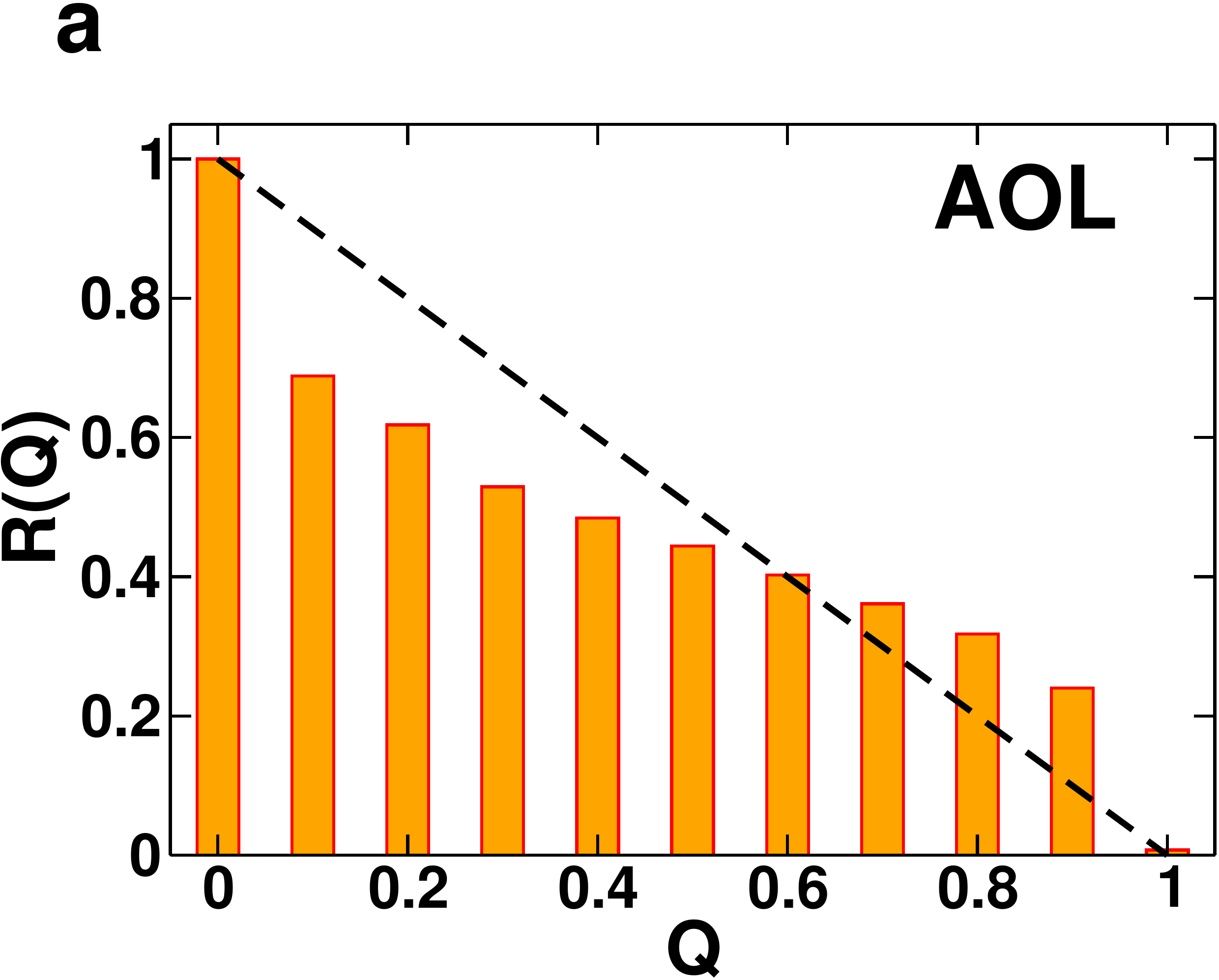}
\qquad
\includegraphics[width=0.3\textwidth, height=0.23\textwidth]{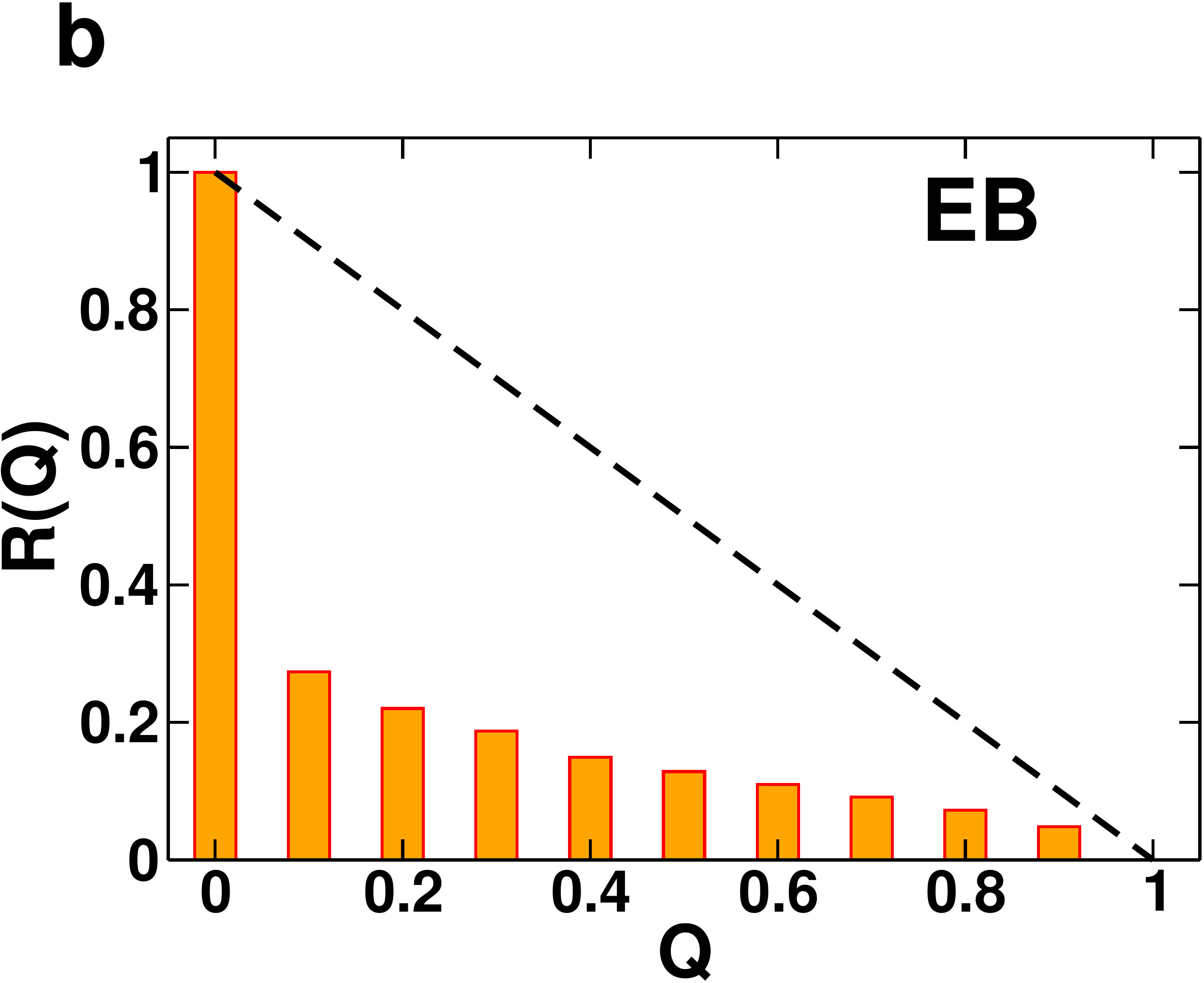}
\qquad
\includegraphics[width=0.3\textwidth, height=0.23\textwidth]{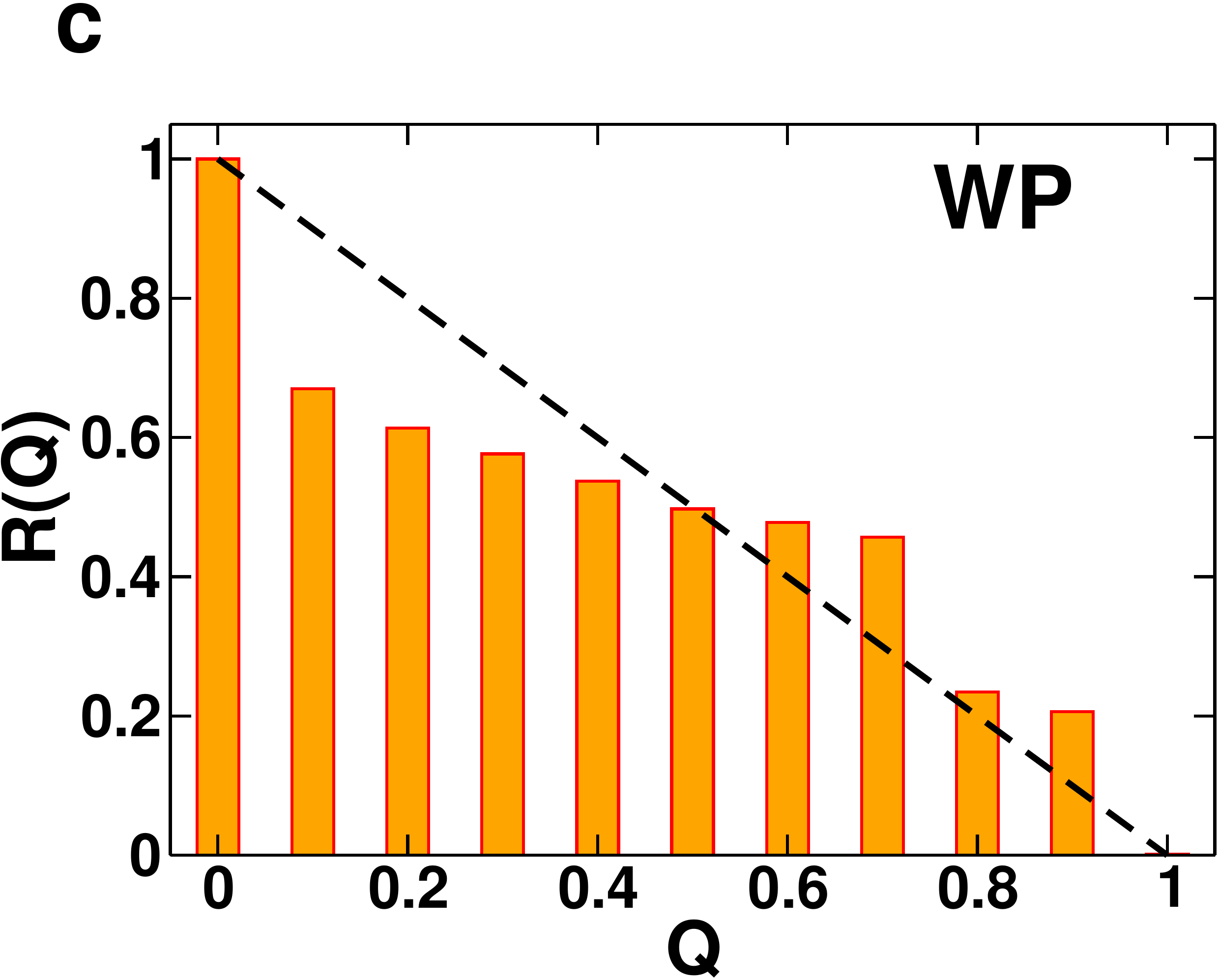}
\caption{(Color online) We report the fraction of users $R\left(Q\right)$ whose inter-event time pdf is described by the the global pdf of rescaled inter-event times $\tilde{P}\left( \tau / \langle \tau \rangle \right)$ with a significance level larger or equal to $Q$. In all cases, the agreement with the theoretical expectation (i.e., dashed lines) is improved if compared with what has been obtained for the global pdf of the unscaled variables (see Fig.~\ref{figks}).}
\label{figks_res}
\end{figure*}

The former analysis has evidenced that the global pdf $P\left(\tau\right)$ is not representative
for the activity patterns of single users. $P\left(\tau\right)$
is measured by averaging 
single users inter-event time pdfs, but such average
is weighted by the pdf of the users's activity.
Since the shape of each $P^{(n)}\left(\tau\right)$
is different, the resulting $P\left(\tau\right)$ represents
therefore an hybrid pdf. This does not necessarily mean
that the behaviors of single users are different, but only that
the assumption that all $\tau$s are drawn from the same underlying distribution
is unlikely.
\\ 
The differences between the  $P^{(n)}\left(\tau\right)$s may 
depend on finite-size effects:
the power-law decay is modulated by periodic
oscillations and additionally may be affected by 
an exponential cutoff. For example, the difference in the decay
exponents, measured in Fig.~\ref{figks2}a, may simply depend on the different range
in which each of these functions is defined
(i.e., the same range in which the power-law fit is performed)
 and the former analysis cannot be considered conclusive.
\\
In this section, we perform an additional statistical test. Instead of considering
the bare value for the inter-event time $\tau$, we take into account
the activity of each single user
and consider the rescaled variable $\tau/\langle \tau \rangle$. $\langle \tau \rangle$
represents the average inter-event time between two actions performed by the same user.
The rescaled variable measures therefore the time gap between
two consecutive operations relative to the typical
(i.e., the average) inter-event time of the single user. This approach
has been already applied in the study of other social systems: 
e-mail~\cite{goh08} and
mobile phone~\cite{candia08} communication systems, 
election~\cite{fortunato07} and citation~\cite{radicchi08} analysis.
In all these papers, it is observed that the scaled variables obey a universal principle
differently from the unscaled variables which generally follow different behaviors.
It should be noticed that the same results may be obtained by considering
$a^{-1}$ (i.e., the inverse of the activity)
instead of $\langle \tau \rangle$ since they are basically the same quantity and
qualitatively similar results may be obtained by considering $n^{-1}$ (i.e., the inverse
of the total number of operations performed) instead of
 $\langle \tau \rangle$ since these quantities are linearly correlated (see Fig.s~\ref{fig2}).
\\
Interestingly, even in the case of our databases, the simple scaling allows
to find a nice collapse between curves corresponding to populations with different
total number of operations. In Fig.~\ref{fig:scale}, for example we plot the quantity $\langle \tau \rangle_n P^{(n)}\left(\tau\right)$ {\it versus} $\tau/\langle \tau \rangle_n$ for the same curves appearing in Fig.~\ref{figks2}a. $\langle \tau \rangle_n = \sum_\tau \, \tau \, P^{(n)}\left(\tau\right)$ stands for the average inter-event time of the whole population of users who have
performed $n$ total operations. 
\\
Even more interestingly, we find that the
global pdf $\tilde{P}\left(\tau/\langle \tau \rangle \right)$ can much
better represents the activity of single users. We perform a KS test as in the former case,
but considering now the scaled variable $\tau/\langle \tau \rangle$ instead of 
$\tau$. The results of this analysis are reported in Fig.s~\ref{figks_res}. Clearly we see
that the relative number of users whose activity pattern is represented by
the global $\tilde{P}\left(\tau/\langle \tau \rangle \right)$ with a significance
level larger or equal to $Q$ is very close to the expected value.
The reliability of $\tilde{P}\left(\tau/\langle \tau \rangle \right)$ is much
higher than the one found for $P\left(\tau \right)$: the percentage of users 
whose activity pattern is represented by the global scaled pdf with
a significance level larger or equal to $Q=0.5$ are $44\%$, $13\%$ and $50\%$
for AOL, EB and WP, respectively, and those values should be compared with
the much worst results, $37\%$ , $5\%$ and $56\%$, obtained in the case of the unscaled pdf.

\section{Waiting time statistics} 
\label{sec:wait}

\begin{figure*}
\includegraphics[width=0.45\textwidth, height=0.345\textwidth]{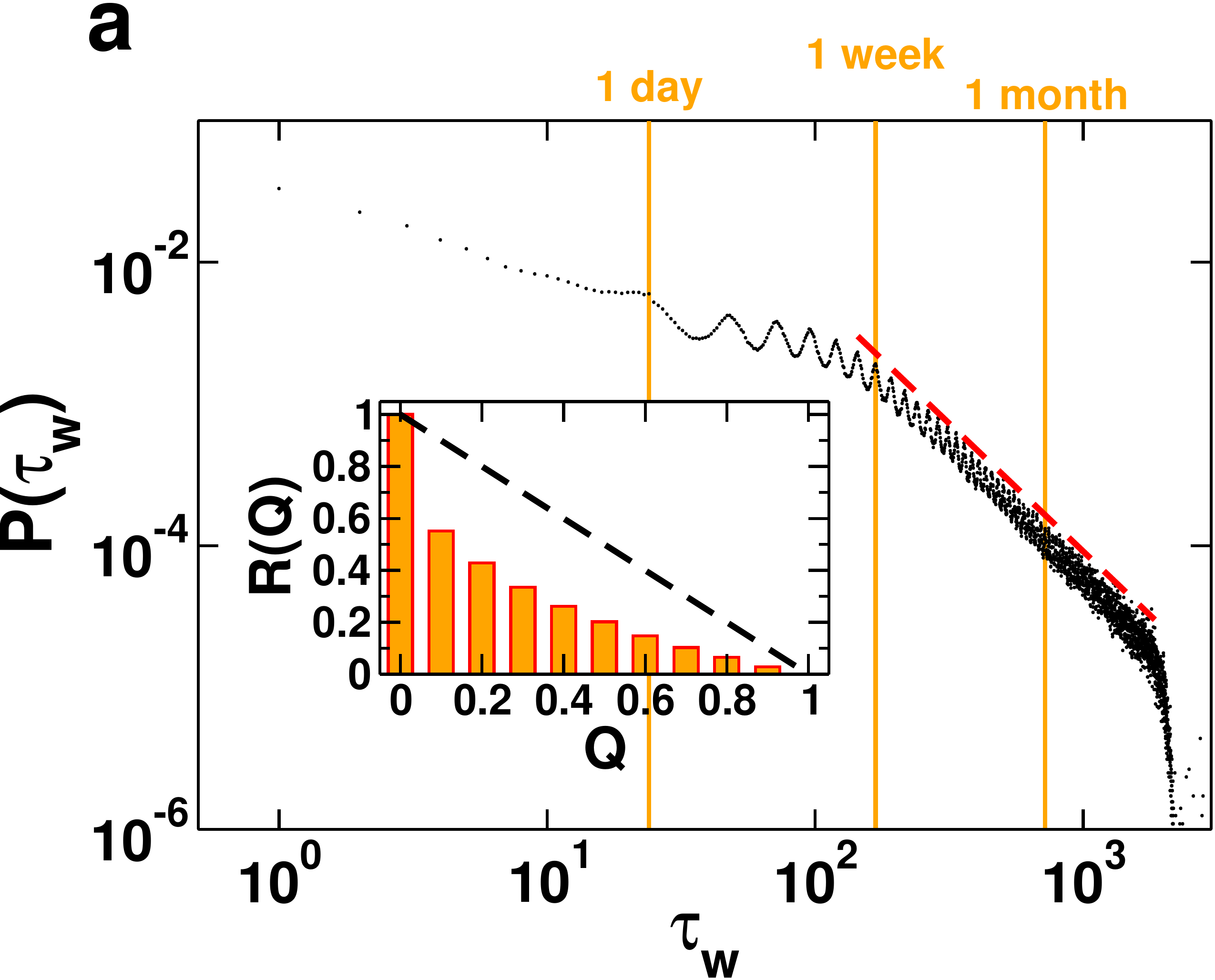}
\qquad
\includegraphics[width=0.45\textwidth, height=0.345\textwidth]{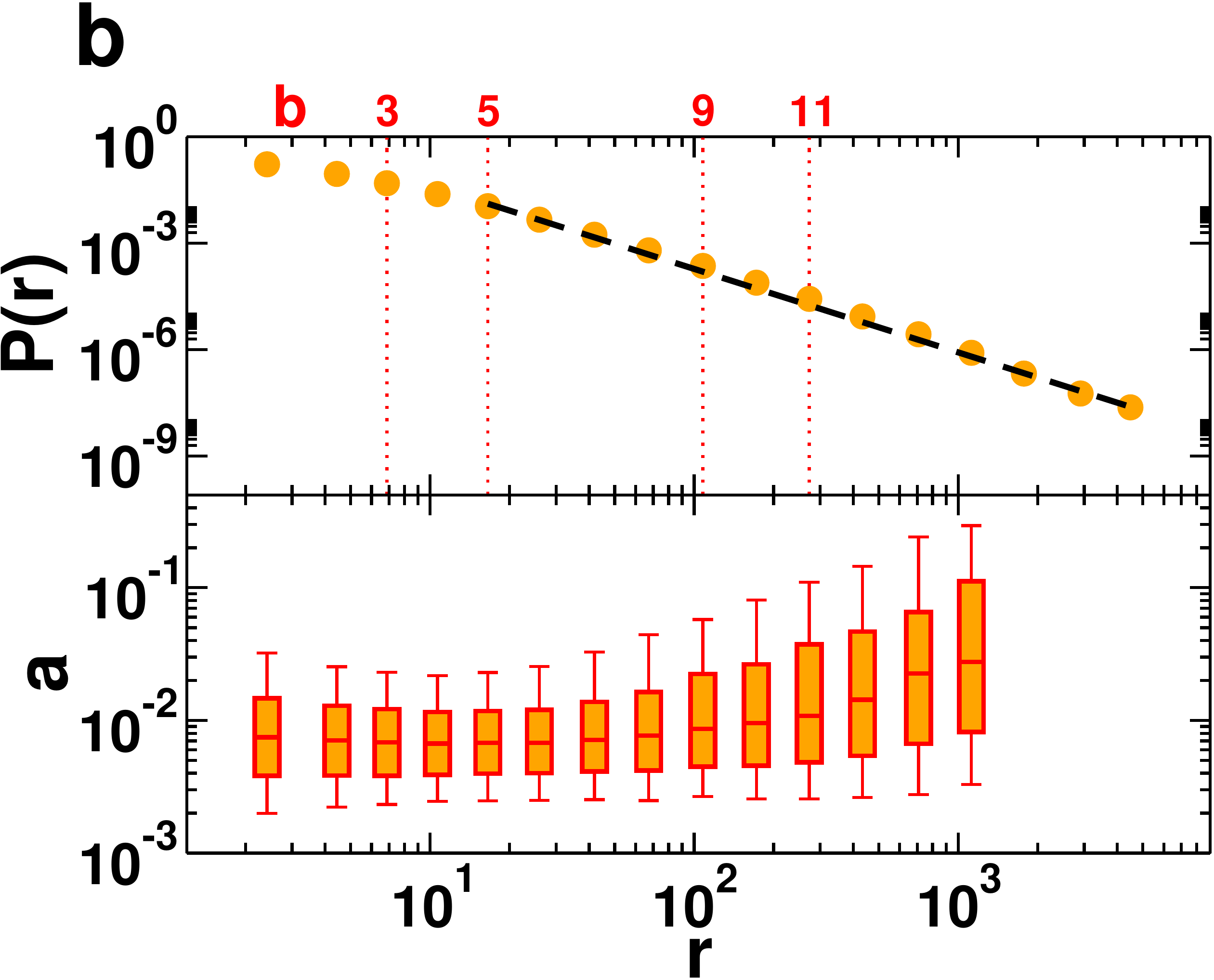}
\\
\vskip 0.2cm
\includegraphics[width=0.45\textwidth, height=0.345\textwidth]{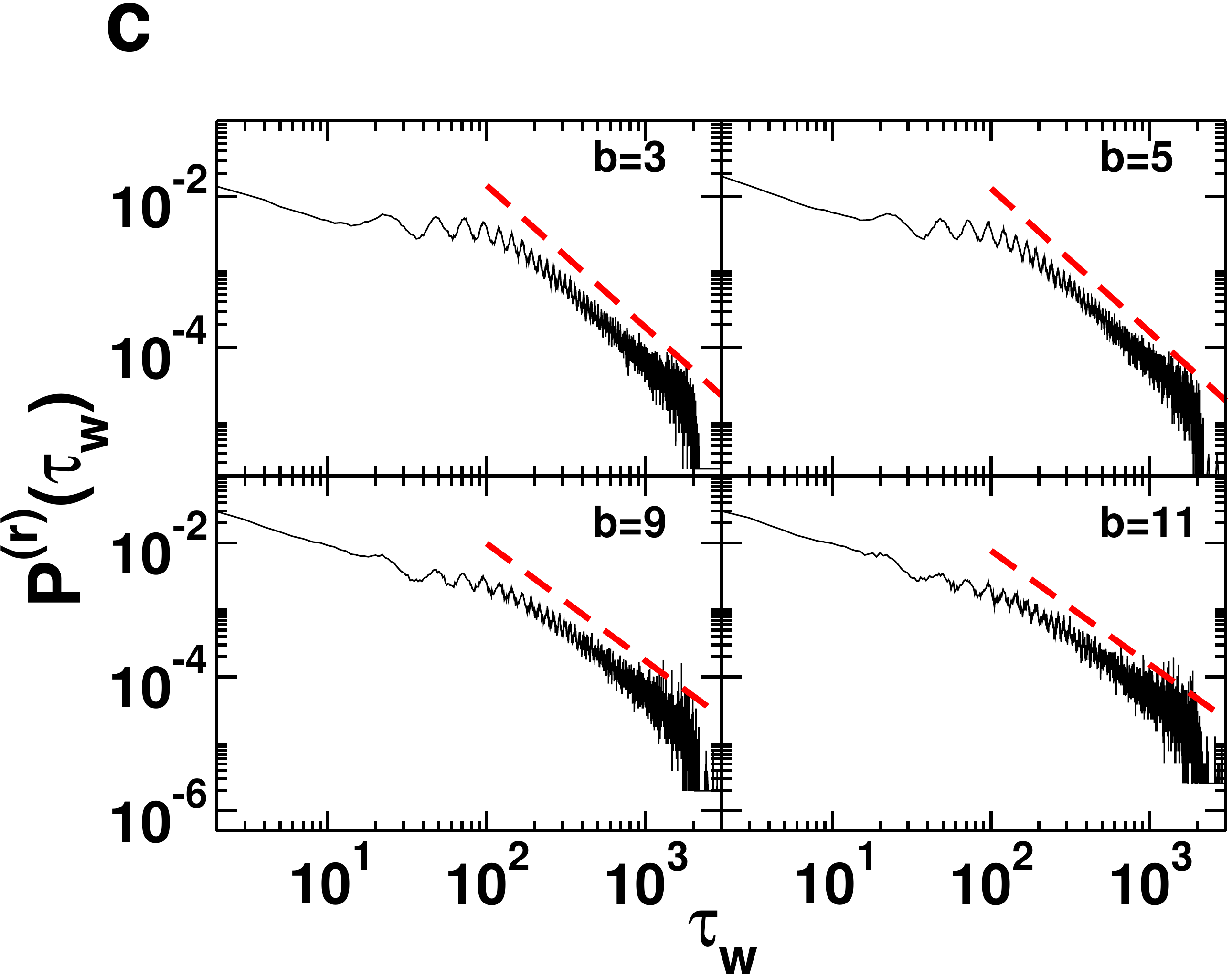}
\qquad
\includegraphics[width=0.45\textwidth, height=0.345\textwidth]{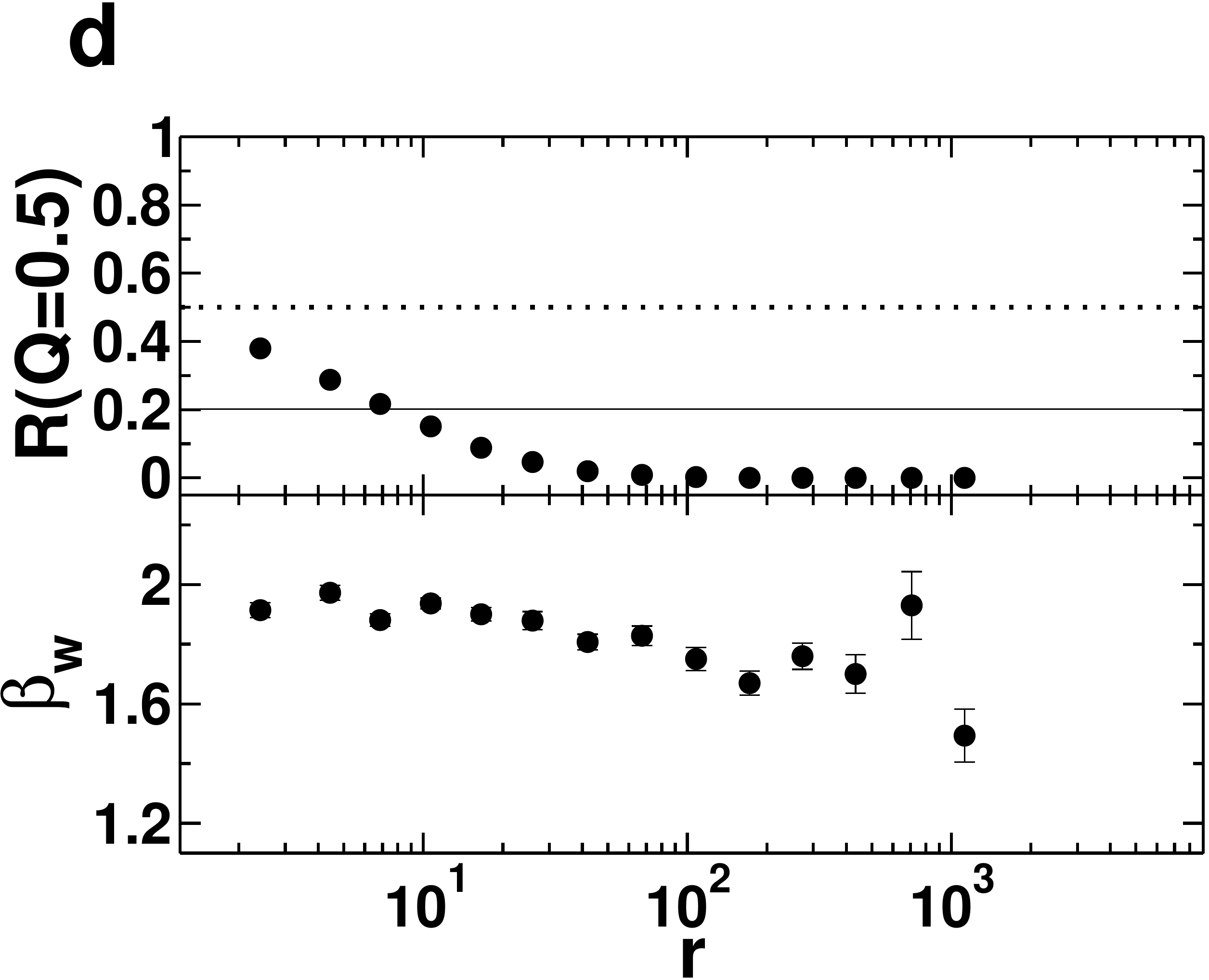}
\caption{(Color online) (a) In the main graph, we plot the global waiting time pdf $P\left(\tau_w\right)$ for EB. The curve is characterized by periodic oscillations and a power-law decay with exponent $\beta_w \simeq 1.8$ (dashed line). In the inset, we report the ratio of users $R\left(Q\right)$ whose replying activity is described by $P\left(\tau_w\right)$ with a level of accuracy at least equal to $Q$. The dashed line represent the theoretically expected behavior of $R\left(Q\right)$. (b) In the top panel, we report the ratio $P(r)$ of users who have sent $r$ replies. The distribution follows a power-law decay with exponent $\lambda_w \simeq 2.3$ (dashed line). In the bottom panel, the average number of replies per unit of time (hour) is plotted as a function of the total number of replies. Error bars denote the values of $a$ corresponding to the top $10\%$ and $90\%$ of each bin. Boxes stand for the values of $a$ referring to the top $25\%$ and $75\%$ of each bin and the horizontal bars corresponds to the median value of $a$ in each bin. (c) Waiting time pdf $P^{\left(r\right)}\left(\tau_w\right)$ corresponding to users who have sent $r$ total replies. Each panel stands for a different bin of those defined in (b): $b=3, 5, 9$ and $11$ which represents users with average number of replies $\langle r \rangle = 6.9, 16.5, 107.9$ and $272.9$, respectively. In all cases we observe a power-law decay and the decay exponents (represented by the slopes of the dashed lines) are: $\beta_w \simeq 1.88, 1.9, 1.75$ and $1.76$. (d) In the top panel, we plot the ratio of users whose waiting time pdf is described by $P^{\left(r\right)}\left(\tau_w\right)$ with an accuracy at least of $Q=0.5$. $R\left(Q=0.5\right)$ is plotted as a function of $r$. The dashed line is the expected value of $R\left(Q=0.5\right)$, equal to $0.5$ in this case. The solid line differently stands for the value of $R\left(Q=0.5\right)$ calculated for the global pdf [see inset of (a)]. In the bottom panel, we plot the decay exponent $\beta_w$ as a function of $r$. In the bottom panel of (b) and in (d) only bins populated by at least $100$ users are shown.}
\label{fig_w}
\end{figure*}

The EB dataset, differently from those of AOL and WP, allows to perform an additional analysis. As already described in section~\ref{sec:data}, all feedback messages we collected from EB contain the ID of the object to which they refer. This information allows to exactly identify feedback messages and their replies. The database offer an error-free source of information to study waiting time pdfs, differently from e-mail datasets where messages and replies can be identified only with heuristics methods~\cite{barabasi05} which can be easily criticized~\cite{stouffer06}.
\\
Consider an object with ID equal to $k$ which has been exchanged during a transaction between the buyer $j$ and the seller $i$. We can compute the reaction time of $i$ to the message sent by $j$ by simply computing the time difference between $t_i^{\left(k\right)}$ and  $t_j^{\left(k\right)}$, which respectively stand for the instants of time when $i$ wrote a feedback message to $j$ and {\it vice versa}. The reply time associated with the object $k$ is therefore given by $\tau_w^{\left(k\right)} = t_i^{\left(k\right)} - t_j^{\left(k\right)}$~\footnote{It should be noticed that the users of EB may live in different part of the world corresponding to different time zones. The knowledge of the country of registration allowed us to report all time stamps to same time zone. In the case of countries, such as US or India for example, in which multiple time zones may coexist, time stamps may be affected by an error of $4$-$5$ hours.}. In our dataset, we are able to find $6\,511\,710$ pairs message/reply which involve $530\,517$ total users. These data are of course a subset of the whole set of data previously analyzed.
\\
In Fig.~\ref{fig_w}a we plot the global waiting time pdf $P\left(\tau_w\right)$. Again, as in the case of the inter-event time pdf, we observe a power-law decay, modulated by periodic oscillations. The decay exponent in this case is $\beta_w \simeq 1.8$. We then perform a KS test in order to estimate the degree of compatibility between the global pdf  $P\left(\tau_w\right)$ and each of the single user's pdf. The results of the KS test are shown in the inset of  Fig.~\ref{fig_w}a: we see that $P\left(\tau_w\right)$ well represents the waiting time pdf of the single users since $R\left(Q\right)$ is reasonable large for each value of the significance level $Q$: for example, the $20\%$ of users have $Q \geq 0.5$. The result is very interesting especially because the values of $R\left(Q\right)$ are much larger than those obtained for the same dataset but in the case of inter-event time statistics (see Fig.~\ref{figks}b).
\\
Also in this case, users show a large heterogeneity in the number of replies they sent. In the top panel of Fig.~\ref{fig_w}b, we plot the relative number of users, namely $P\left( r \right)$, who have sent $r$ reply messages. $P\left( r \right)$ decays power-like as $r$ increases [i.e., $P\left( r\right) \sim r^{-\lambda_w}$] with exponent  $\lambda_w \simeq 2.3$. However, to the heterogeneity in the number of replies does not correspond an heterogeneity in the activity. The average number of replies sent in a unit of time $a$ is plotted in the bottom panel of Fig.~\ref{fig_w}b: $a$ does not strictly depend on $r$, since its value is almost constant for all $r$ and shows only a slight increase for large values of $r$.
\\
The homogeneity in $a$ is reflected in the waiting time pdfs $P^{\left(r\right)}\left(\tau_w\right)$, relative to users who have sent $r$ total replies. In Fig.~\ref{fig_w}c, we plot $P^{\left(r\right)}\left(\tau_w\right)$ calculated for subsets of users who have sent a similar number of replies. For simplicity, we consider the same division in bins as defined in both plots of Fig.~\ref{fig_w}b. As we can see, independently of the value of $r$ the waiting time pdfs decay power-like [i.e., $P^{\left(r\right)}\left(\tau_w\right) \sim \tau_w^{-\beta_w}$] as a function of $r$ and the decay exponent  is always close to $1.8$. The same is true also for other values of $r$: in the bottom panel of Fig.~\ref{fig_w}d, we plot the decay exponent $\beta_w$ as a function of $r$ and we can clearly see that $\beta_w$ is almost the same in all cases. As final result, in the top panel of Fig.~\ref{fig_w}d, we consider the ratio of users, with total number of replies $r$, whose waiting time pdf is identical to $P^{\left(r\right)}\left(\tau_w\right)$ with a probability $Q \geq 0.5$: as in the case of inter-event time distributions, also in this case the degree of compatibility decreases suddenly to zero as $r$ increases.

\section{Conclusions}
\label{sec:end}

In this paper, we have studied some statistical properties of human activities in the Web. We have analyzed three completely different systems:  search's inquires performed in the search engine of America On Line (AOL), feedback messages exchanged by users of Ebay (EB) and logging actions of users in the English website of Wikipedia (WP). These systems are clearly different each other for various reasons. The main difference is given by the range of interaction between users: in AOL, users are totally independent; in EB, communications are restricted between two users; in WP each user's action is dependent on the actions performed by a group of other users. Despite this difference, the global emergent behavior is very similar: $P\left(\tau\right)$, which is the relative number of subsequent human actions which differ by an amount of time $\tau$, decreases power-like as $\tau$ increases. The bursty behavior seems therefore to be intrinsic to human nature and not due to the interaction (and the type of interaction) with other humans.
\\
However, the global inter-event time probability distribution function (pdf) $P\left(\tau\right)$ is not well representative for the behavior of single users. The single user's pdf of the absolute inter-event time is dependent on how much the user is active. We have restricted the calculation of the inter-event time pdfs only to users with the same number of operations $n$, namely $P^{\left(n\right)}\left(\tau\right)$, and we have found, by using a statistical non-parametric test, that each $P^{\left(n\right)}\left(\tau\right)$ represents its corresponding population very well. The degree of compatibility of each $P^{\left(n\right)}\left(\tau\right)$ is in general much better than that one of the global each $P\left(\tau\right)$.
This fact, already noticed in other systems~\cite{zhou08}, has deep consequences. If one measures the global pdf of the bare inter-event time, the resulting function
is a weighted superposition of apparently different pdfs defined over clearly different ranges.
In this sense, the poor reliability of  $P\left(\tau\right)$ is due not to an intrinsic different behavior of the users,
but to the wrong way to observe the system. We have however found the way to pass over this obstacle.
Instead of considering the pure values of the inter-event times, one should suppress the
observed dependence on the activity and consider relative quantities. By replacing $\tau$ with 
$\tau / \langle \tau \rangle$, all users can be  compared in a fair way and the resulting
pdfs (single users'ones and the global one) are significantly equivalent.

\

We have finally studied  the waiting time pdf in EB communications. We have performed the same kind of analysis conducted in the case of inter-event time pdfs, but we have found an interesting difference. Despite users are heterogeneous in the number of replies, their average number of replies per unit of time is almost the same. The consequence is that all waiting time pdfs  $P^{\left(r\right)}\left(\tau_w\right)$, corresponding to users who have sent $r$ total replies, are almost identical and their decay exponents are compatible with the one of the global pdf $P\left(\tau_w\right)$. 

\

In conclusion, spontaneous activity seems to do not obey any universal rule if
one observes the system on an absolute scale. The inter-event time pdfs of single users decay power-like 
with exponents ``apparently'' dependent on how much the users are active.
This is however due to the wrong way to monitor the system. The spontaneous activity of each single user
is triggered by her/his own internal ``biological'' clock. Inter-event times should therefore weighted
on different scales by using different units of measure. When absolute quantities are replaced
by relative ones, the apparently different behavior becomes more similar and a universal rule
governing the activity of humans in the Web emerges.
In future investigations, inter-event time pdfs should be studied by taking
this fact into account.  On the other hand, the time patterns of replying activities seem to be coherent among users.
People seems to react to external stimuli in the same identical way.
Further investigations are needed in this direction and the analysis of other communication databases might provide evidence to the results showed in this paper.

\begin{acknowledgments}
Thanks to S.~Fortunato, R.D.~Malmgren, A.~Lancichinetti and J.J.~Ramasco for useful comments and suggestions.
We thank also an anonymous referee for a constructive criticism which has become an important 
suggestion for the improvement of the quality of the paper.
\end{acknowledgments}

\end{document}